\documentclass[12pt]{iopart}

\usepackage{graphicx}
\usepackage{dcolumn}
\usepackage{bm}
\begin{document}
\newcommand{\eqb}{\begin{eqnarray}}
\newcommand{\eqe}{\end{eqnarray}}
\newcommand{\diff}{\textrm{d}}
\newcommand{\Eclassical}{E_{\rm cl}}
\newcommand{\Ecrit}{E_{\rm crit}}
\newcommand{\alphafine}{\alpha_{\rm f}}
\newcommand{\ehatvec}{\hat{{\bm e}}}
\newcommand{\betavec}{\bm{\beta}}
\newcommand{\lambdamicron}{\lambda_{\mu}}
\newcommand{\strength}{\hbox{$a$}}
\newcommand{\wsqcm}{\textrm{W\,cm}^{-2}}
\newcommand{\melec}{m_{\rm e}}
\newcommand{\eps}{\Delta}

\title{Pair production in counter-propagating laser beams}

\author{J G Kirk,$^1$ A R Bell$^{2,3}$ and I Arka$^1$}

\address{$^1$ Max-Planck-Institut f\"ur Kernphysik, 
Postfach 10 39 80, 69029 Heidelberg, Germany}
\address{$^2$ Clarendon Laboratory, University of Oxford, 
Parks Road, Oxford OX1 3PU, UK}
\address{$^3$ STFC Central Laser Facility, RAL, Didcot OX11 0QX, UK}
\ead{john.kirk@mpi-hd.mpg.de}
\begin{abstract}
Based on an analysis of a specific electron trajectory
in counter-propagating beams, 
Bell \& Kirk (PRL 101, 200403 (2008)) 
recently suggested that laboratory lasers
may shortly be able to produce significant numbers of 
electron-positron pairs. 
We confirm their results using an improved treatment of 
nonlinear Compton scattering in 
the laser beams. Implementing an algorithm that 
integrates classical electron trajectories, we then 
examine a wide range of laser pulse shapes and polarizations. We find that 
counter-propagating, linearly polarized 
beams, with either aligned or crossed orientation, are likely to initiate 
a pair avalanche at intensities of approximately $10^{24}\,\wsqcm$ per beam. 
The same result is found by modelling one of the beams
as a wave reflected at the surface of an overdense solid. 
\end{abstract}

\submitto{\PPCF}

\section{Introduction}
Within the next few years, powerful lasers may be able to realise
intensities of $10^{23}$ to $10^{24}\,\wsqcm$ in the laboratory,
enabling novel physical processes to be investigated
\cite{kogaetal06,muelleretal08,dipiazzahatsagortsyan08}. 
One of these is the prolific
production of electron-positron pairs predicted to occur when an
electron is accelerated in counter-propagating laser beams \cite{bellkirk08}.

Pair production has already been realised in the lab.\ using 
laser intensities $\sim10^{20}\,\wsqcm$. The technique, proposed 
in \cite{sheareretal73},
involves 
using the laser beams to accelerate electrons to MeV energies, and then
allowing them to produce an electromagnetic cascade in a foil made of 
high-$Z$ target material. Depending on the experimental set-up,
positrons are created either by the trident process, 
in which an energetic electron interacts with the electrostatic field
of a target nucleus, or via
an intermediate gamma-ray (produced by bremsstrahlung)
that subsequently pair-creates in the field of a nucleus by the 
Bethe-Heitler process 
\cite{liangetal98,cowanetal99,nakashimaetal02,chenetal09}.
The technique operates at relatively modest laser intensity, but
converts only a small fraction of the laser pulse energy into pairs.  

An alternative mechanism that has been intensively studied but not yet
observed
is the spontaneous creation of pairs out of the vacuum
by laser beams. Schwinger \cite{schwinger51} 
predicted spontaneous pair creation 
in an electric field that approaches the critical
value $E_{\rm crit}=1.3\times10^{18}\,\textrm{V\,m}^{-1}$, which is
achieved at a laser intensity of roughly $10^{29}\,\wsqcm$.
Conservation of energy and momentum forbids this process in a
plane-wave beam, but it can operate in counter-propagating beams
\cite{bulanovetal06,rufetal09} and may carry interesting information
on the beam sub-structure \cite{hebenstreitetal09}.
The effect does not have a sharp threshold, but the rate 
drops rapidly as the amplitude $E$ of the electric field decreases, since it
contains a factor $\textrm{exp}\left(-E_{\rm crit}/E\right)$. 
Detailed calculations predict observable consequences at intensities as low as
$10^{26}\,\wsqcm$ \cite{bulanovetal06}, but these are unlikely to be
achieved within the next few years. 

A third mechanism of pair production using an intense laser beam was
realised in an experiment at SLAC \cite{burkeetal97}. A beam of
$46.6\,$GeV electrons was fired into a laser beam, where electron
positron pairs were created. This was interpreted as being due to a
two-step process: First a GeV photon was created by nonlinear Compton
scattering of multiple laser photons by a relativistic electron. Then
this photon interacted with multiple laser photons to create a
pair. This technique is basically the same as the first method described
above, except that the electrostatic field of the target nucleus is
replaced by the electromagnetic field of the laser. The possibility
that pairs might be directly produced by the trident process in the
laser fields was not discussed. The positron yield observed in this
experiment was very small.

The mechanism for {\em prolific} pair production suggested by
\cite{bellkirk08} is closely related to that realised at
SLAC. However, instead of using a particle accelerator, the laser
beams themselves accelerate the relativistic electrons. The
entire cycle of electron acceleration and pair creation (either by the
electromagnetic analogue of the trident process or via an intermediate
real photon) takes place when laser beams counter-propagate in an
under-dense plasma. At intensities $\sim10^{24}\,\wsqcm$ the number of
pairs created per plasma electron was estimated by \cite{bellkirk08}
to be about unity. Since each new electron and positron is, in its
turn, accelerated in the laser beams, one expects an avalanche of
pairs that should absorb a significant fraction of the laser pulse
energy.  This process would, therefore, swamp the pure vacuum effect.

The prediction of \cite{bellkirk08} was based on the analysis of a
particularly simple electron orbit at the magnetic node in the field
of two counter-propagating, circularly polarized pulses, and used
simplified descriptions of the physical processes. In particular, the
photon emissivity was treated using a monochromatic approximation.  In
this paper we examine this mechanism more closely.  The analysis is
extended in two ways: we use improved approximations to the physical
processes, and embed these in a numerical scheme that follows the
trajectory of an electron in more realistic models of the laser
fields. Vacuum fields are used to describe the laser beams, which is
appropriate if they propagate in an underdense plasma, and we examine the
trajectories of initially nonrelativistic electrons picked up by the beams at
different points in the plasma. 

In section~\ref{processes} we give details of the relevant physical
processes. These all involve multi-photon interactions with the 
two laser beams. However, we show that, 
in the parameter range of interest, 
they can also be viewed as interactions of an electron 
or high-energy photon with {\em uniform, static} 
electromagnetic fields. 
The processes are (a) photon emission by a
relativistic particle. In various special cases this is known as synchrotron
radiation, magneto-bremsstrahlung, curvature radiation, non-linear
Thomson scattering or non-linear Compton scattering; here we use the
term \lq\lq synchrotron radiation\rq\rq\ to mean the generic
process. We will assume the static field is incorporated exactly
into the electron propagator, in which case the process is of first
order in the fine-structure constant $\alphafine$. 
(b) The
second-order (in $\alphafine$) process of pair production by a charged
particle in a static field. (This involves an
intermediate virtual photon and is the electromagnetic analogue 
of the trident process. In the following we simply refer to it as the 
\lq\lq trident process\rq\rq.) (c) The first-order process of pair
production by a real photon (a synchrotron photon, for example) in
the same static field.

In section~\ref{trajectories}
we describe our model of the laser fields, and give the classical
equations of
motion of the electron including radiation reaction.  The way in which
pair creation by the processes described in section~\ref{processes} is
implemented is discussed in section~\ref{paircreation}.  
Our results fall naturally into two parts. In 
section~\ref{longpulses} we present calculations
in counter-propagating, circularly polarized pulses with the same sense of 
rotation of the fields. The motivation here is to compare our results with
\cite{bellkirk08} and to understand the influence of the 
improved treatment of the photon emissivity and of different choices 
of initial conditions of the electron trajectory. In section~\ref{shortpulses}
we then examine pair production in 
more realistic pulses, with both circular and linear polarization.
These pulses have a finite duration and the fields are assumed to be 
contained within a cylindrical region whose axis lies along the 
propagation direction. Our conclusions, which confirm and extend the 
predictions made in \cite{bellkirk08}, are summarized
in section~\ref{conclusions}.

\section{Physical processes}
\label{processes}
\subsection{The quasi-stationarity and weak-field approximations}
For an electron in a monochromatic plane wave, the photon emissivity
and pair-creation rate is a function of only two dimensionless,
Lorentz invariant parameters \cite{ritus79}.  These are the strength
parameter of the wave \eqb a&=&\frac{e E_0}{m c \omega_{\rm laser}}
\eqe where $E_0$ is the amplitude of the electric field and
$\omega_{\rm laser}$ its angular frequency, and the parameter $\eta$
that determines the importance of strong-field quantum effects: 
\eqb
\eta&=&\frac{e\hbar}{m^3c^4} \left|F_{\mu\nu}p^\nu\right|
\label{etadef}
\eqe
where $p^\mu$ is the four-momentum of the electron, $F^{\mu\nu}$ is the 
electromagnetic field tensor, and  
$\left|\dots\right|$ denotes the length of the four-vector.  
In terms of the field components $E_i,B_i$
in Cartesian coordinates, with $i=1,\dots3$:
\eqb
F^{i0}&=&E_i\qquad F^{ij}\,=\,-\varepsilon_{ijk}B_k\qquad
F^{\mu\nu}\,=\,-F^{\nu\mu}
\nonumber
\eqe
 and $\varepsilon_{ijk}$ is the antisymmetric Levi-Civita symbol.

The interaction of a particle with an arbitrary external
electromagnetic field is more complex.  However, two approximations
simplify it considerably. The first is that of quasi-stationarity:
\lq\lq instantaneous\rq\rq\ values of the transition rates are
computed assuming the external field is constant in time, and these
values are subsequently time-averaged.  This requires that the
variation timescale of the field is long compared to the coherence
time associated with the interaction.  In the monochromatic,
plane-wave case, the coherence time is 
$t_{\rm coh}\approx\left(E_{\rm crit}/E_0\right)\left(\hbar/mc^2\right)$~\cite{ritus79}.  Therefore,
the variation timescale of the field is long compared to the coherence
time of the interaction provided
\eqb a\gg1
\label{strengthcondition}
\eqe 
If this (Lorentz invariant) condition is satisfied, the instantaneous
transition probabilities for an electron of given $p^\mu$ can be
computed in the limit $a\rightarrow\infty$, i.e., for uniform, static
fields. 
We assume that this argument can be
generalized to the case of counter-propagating waves
each of strength parameter $a$. A potential  
problem arises for those configurations where there exist 
points in space that are simultaneously 
nodes of both the electric and magnetic fields, since there the 
coherence length of the interaction becomes large. However, the processes
that are important for pair production are confined to the regions of 
strong field, where the coherence length is short.  

The situation can readily be visualized in the classical picture:
the coherence length associated with the 
synchrotron radiation of a 
relativistic electron of Lorentz factor $\gamma$, 
is the path length over which the electron is deflected
by an angle $1/\gamma$, which gives the same 
result as in the quantum case:
$t_{\rm coh}\approx mc/\left(\left|e\right|E_0\right)$. 
Variations of the accelerating fields
on lengthscales shorter than the deflection length 
occur only close to simultaneous nodes of the $\bm{E}$ and $\bm{B}$
fields. However, only relatively low frequency
radiation is emitted in these regions.

The second approximation is that of weak fields. In an arbitrary,
constant field, the transition probabilities may depend not only on
$\eta$, but also on the two Lorentz invariant parameters associated
with the field: $f=\left|E^2-B^2\right|/E_{\rm crit}^2$ and
$g=\left|\bm{E}\cdot\bm{B}\right|/E_{\rm crit}^2$.  Provided, however, that 
\eqb
f\ll 1\ \textrm{and}\ g\ll 1
\label{weakfield}
\eqe
and 
\eqb
\eta^2\gg \textrm{Max}(f,g)
\label{weakcondition}
\eqe one may neglect this dependence and evaluate the transition
probabilities in any convenient field configuration that has the same
value of the parameter $\eta$ \cite{baierkatkov68,ritus79}.  Analogously,
pair production by a single high-energy
photon in a laser field with $a\gg1$ can be approximated as pair production
in a uniform, static field, and the transition probability 
for interaction with the virtual photons of this field
depends only on the parameter 
\eqb
\chi&=&\frac{e\hbar^2}{2m^3c^4}\left|F^{\mu\nu}k_\nu\right|
\label{chidef}
\eqe
where $\hbar k^\mu=(\hbar\omega,\hbar\bm{k})$ is the photon four-momentum.

The situations we consider in this paper fulfil both the
quasi-stationary requirement (\ref{strengthcondition}) and the
weak-field conditions (\ref{weakfield}) and (\ref{weakcondition}). At
a laser wavelength of $\lambda_{\mu{\rm m}}\,\mu\textrm{m}$, the
strength parameter in a single, linearly polarized beam of intensity
$I_{24}\times 10^{24}\,\wsqcm$ is $a=855 I_{24}^{1/2}\lambda_{\mu{\rm
    m}}$, so that the stationarity condition~(\ref{strengthcondition}), 
is well satisfied in the range of
interest ($\lambda_{\mu{\rm m}}=1$, $I_{24}=0.1\dots 1$). It is
possible to configure laser beams in such a way as to reduce the
variation timescale.  For example, in a
beam that is reflected from an overdense solid, high harmonics are
generated~\cite{baevaetal06,baeva08}.  
However, unless substantial power is present in harmonics
numbers $n>100$, the stationarity approximation remains valid. The
maximum field strength in a monochromatic, linearly polarized beam is
$E/E_{\rm crit}=2.1\times10^{-3}I_{24}^{1/2}$, and the field invariants
$f$ and $g$ in (\ref{weakfield}) vanish in such a wave.
In counter-propagating beams 
(each of intensity $I_{24}\times 10^{24}\,\wsqcm$), 
they do not vanish, but, although
the field reaches twice the amplitude
of the individual beams, one still has at all space-time points
$f,g<1.6\times10^{-5}I_{24}$,
so that (\ref{weakfield}) is satisfied. 
Pair production becomes important for $\eta>0.1$, which requires
relativistic electrons with $\gamma>50\,I_{24}^{-1/2}$.
In this range, the
weak field condition~(\ref{weakcondition}) is easily satisfied.
Consequently, the transition probabilities for photon production and
pair production by an electron (via a virtual photon) and by a real
photon may be taken from computations performed for static,
homogeneous magnetic
fields that have the same value of the parameter $\eta$ or $\chi$,
provided that this static configuration also fulfils the weak-field
conditions.

In particular, choosing a situation in which an electron of Lorentz
factor $\gamma$, or a photon of energy $\epsilon mc^2$,
propagates normal to a constant, uniform magnetic field $B$, the
relevant parameters are $\eta=\gamma B/B_{\rm crit}$ and
$\chi=\left(\epsilon/2\right)B/B_{\rm crit}$.  
Since $f=\left(B/B_{\rm crit}\right)^2$, 
the weak-field condition (\ref{weakcondition})
requires us to compute these transition probabilities for a field $B\ll
B_{\rm crit}$, i.e., in the limit $\gamma,\epsilon\gg1$.  This case
corresponds to the classical limit for the electron trajectory, where
the particle energy is large compared to $\hbar eB/mc$, and to the
case well above the threshold at $\epsilon=2$ for pair creation, in
the case of the photon.  These probabilities were calculated in the
1950's; they are conveniently reviewed by Erber, \cite{erber66}, whose
results we summarize in the following. Their application to arbitrary fields,
as outlined above, 
and their generalization to particles of arbitrary spin 
is discussed by Baier \& Katkov~\cite{baierkatkov68}.

\begin{figure}
 \includegraphics[width=\textwidth]{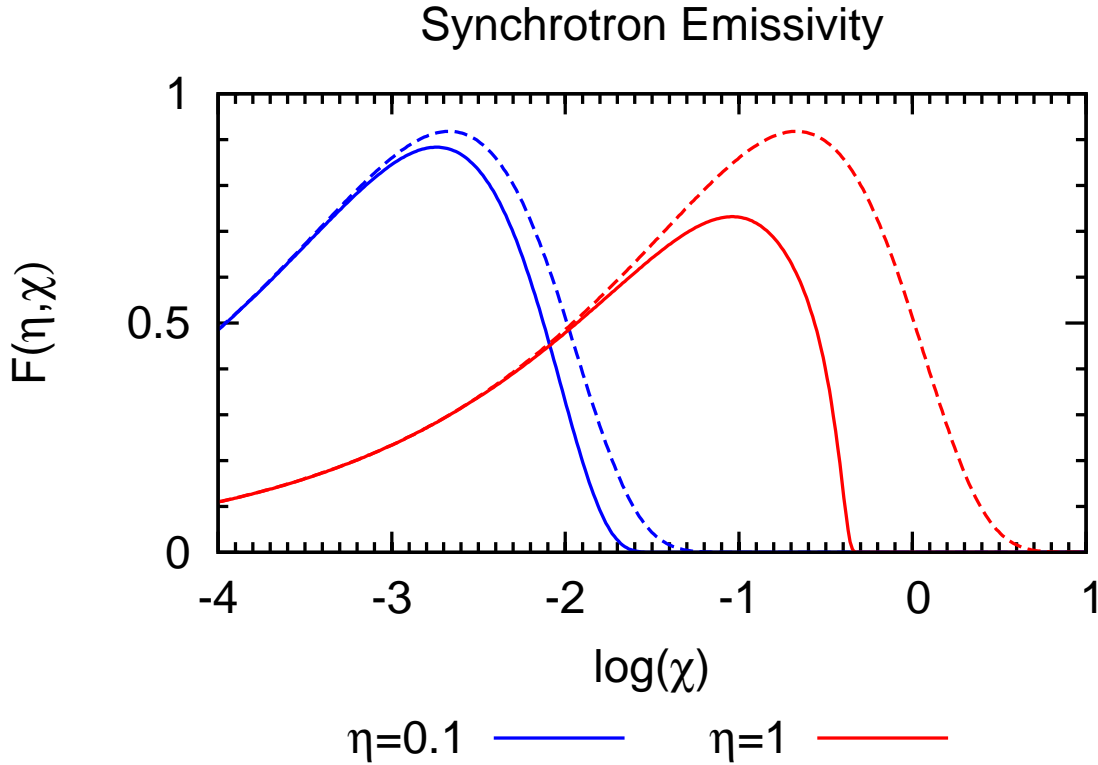}
\caption{\label{synchrotronfig}%
The quantum synchrotron
  function $F(\eta,\chi)$ as a function of the dimensionless photon
  frequency $\chi$ (see (\ref{chidef})). The parameter $\eta$,
  defined in (\ref{etadef}), determines the importance of strong
  field QED effects; the energy of the radiated photon would equal
  that of the electron at $\log\chi=\log(\eta/2)$. The 
corresponding classical functions (see (\ref{synchclassical}))
are shown as dashed lines.
 }
\end{figure}

\subsection{Synchrotron radiation}
The rate of production of photons by an electron of energy $\gamma
mc^2$ moving normal to a magnetic field of strength $B=bB_{\rm crit}$
is defined as 
\eqb 
\frac{\diff^2 N}{\diff\chi\diff
  t}&=&\sqrt{3}\frac{mc^2}{h}\alphafine b \frac{F(\eta,\chi)}{\chi}
\eqe where $\chi$, defined in (\ref{chidef}), describes the energy
of the emitted photon.  In this field configuration, $\eta=\gamma b$,
and we can multiply this equation by $\gamma$ to display the Lorentz
invariance of each side: 
\eqb 
\gamma\frac{\diff^2 N}{\diff \chi\diff
  t}&=&\sqrt{3}\frac{mc^2}{h}\alphafine \eta \frac{F(\eta,\chi)}{\chi}
\label{synchemissivity}
\eqe 
For $\gamma\gg1$ and $b\ll1$, (i.e., under the weak-field
conditions (\ref{weakfield}) and (\ref{weakcondition})) the function
$F(\eta,\chi)$ is given in equation~(2.5a) of \cite{erber66}, and is
reproduced in our notation in \ref{appendix_synchrotron}.  In the
classical limit, it becomes a function of $\eta$ and $\chi$ in the
combination $\chi/\eta^2$: 
\eqb 
F(\eta,\chi)&\rightarrow&
f_{\rm synch}\left[4\chi/\left(3\eta^2\right)\right]
\ \textrm{as}\ \hbar\rightarrow0
\label{synchclassical}
\eqe 
where $f_{\rm synch}$ is the familiar expression for the
synchrotron emissivity (summed over polarizations) in the Airy
integral approximation \cite{melrose80}: 
\eqb 
f_{\rm synch}(y)&=&y\int_y^\infty\,\diff t\, K_{5/3}(t) 
\eqe 
The
monochromatic approximation used in \cite{bellkirk08} is equivalent to
the replacement 
\eqb 
F(\eta,\chi)&\rightarrow&f_{\rm  mono}
\left[4\chi/\left(3\eta^2\right)\right] 
\eqe 
with 
\eqb 
f_{\rm mono}(y)&=&\frac{8\pi}{9\sqrt{3}}\delta\left(y-y_0\right)
\label{fsynchmono}
\eqe
together with the choice $y_0=0.29$ (which is where the function
$f_{\rm synch}(y)$ has its maximum). The coefficients in front of the
$\delta$-function in (\ref{fsynchmono}) ensure that the power
radiated by an electron is the same in the classical and the
monochromatic approximations.

In our problem, the most important difference between the quantum and
classical emissivities is that the quantum emissivity 
is severely depleted for hard photons, 
when $\chi\sim\eta\sim1$, and vanishes for
photon energies larger than that of the incoming electron: 
\eqb
F(\eta,\chi)&=&0\ \textrm{for}\ \chi\ge \eta/2 
\eqe 
This has an important influence on pair
production, since these photons subsequently have the highest
probability of conversion. The effect is illustrated in
figure~\ref{synchrotronfig}. It is associated with a reduction in the
(Lorentz invariant) 
total power $P$ radiated by the electron (integrated over all emitted
photons):
\eqb
P&=&\frac{2}{3}\alphafine\eta^2\,mc^2\frac{mc^2}{\hbar}g(\eta) 
\label{totalpower}
\eqe
with 
\eqb
g(\eta)&=&\frac{3\sqrt{3}}{2\pi\eta^2}\int_0^\infty\,\diff\chi\,F(\eta,\chi)
\label{defg}
\eqe
The classical result is $g(\eta)=1$. For $\eta\ll1$ the lowest order
quantum correction gives \cite{ritus79} \eqb
g(\eta)&\approx&1-\frac{55\sqrt{3}}{16}\eta \eqe The full expression
is given in \ref{appendix_synchrotron}.

\subsection{Trident pair production via virtual photons}
Trident pair production via virtual photons is a second order
process. The rate, computed using a Weizs\"acker-Williams
approximation, is presented by Erber \cite{erber66}. This
approximation involves treating the electromagnetic field of the
incident electron as a superposition of virtual photons. The
computation then essentially evaluates the dispersion relation for
these photons in the magnetized vacuum. The imaginary part of the
refractive index gives the absorption rate, i.e., the rate of pair
production. This procedure should provide a reasonable approximation
when the weak-field conditions are satisfied.

The rate of pair production for an electron of energy $\gamma mc^2$
moving normal to a constant magnetic field $B=bB_{\rm crit}$ is
(\cite{erber66}, equation~(4.4)):
\eqb
\gamma\frac{\diff N_\pm}
{\diff t}&=&0.64\frac{mc^2}{h}\alphafine^2\eta\hat{\Omega}(\eta)
\label{virtualrate}
\eqe
The function $\hat{\Omega}(\eta)$ is given in
\ref{virtualappendix}.
For small $\eta$ it can be approximated by 
\eqb
\hat{\Omega}(\eta)&\approx&
\frac{\pi^{5/2}}{16}\left(3\eta\right)^{1/4}\textrm{exp}
\left[-8/\left(3\eta\right)^{1/2}\right]
\label{tridentapprox}
\eqe
For $\eta\gg1$, Erber gives the asymptotic value
$\hat{\Omega}(\eta)\rightarrow\left(\pi^2/2\right)\ln\eta$.

\subsection{Pair creation by synchrotron photons}

The photon absorption coefficient in the ultra-relativistic limit required by
the weak-field approximation is expressed by 
Erber \cite{erber66} as an absorption probability per unit path 
length. Since we will be interested in the propagation of photons in
electromagnetic fields that vary in space and time, we write the instantaneous
absorption probability in terms of the differential optical depth
$\diff\tau$ traversed by the photon in an interval of time $\diff t$ in the
lab.\ frame:
\eqb
\frac{\diff \tau}{\diff t}&=&
\alphafine\frac{mc^2}{\hbar}\frac{mc^2}{h\nu}
\chi T_\pm(\chi)
\label{photonabsorption}
\eqe
The function $T_\pm(\chi)$ is given approximately by
\eqb
T_\pm(\chi)&\approx&
0.16\frac{K^2_{1/3}\left(\frac{2}{3\chi}\right)}{\chi} 
\label{photonabsorption2}
\eqe
For small $\chi$ it is proportional to
$\textrm{exp}\left[2/\left(-3\chi\right)\right]$ and
so is exponentially small, despite that fact that, 
in the weak-field approximation, the photon is always well
above the kinematic threshold. The function peaks at $\chi\approx8$ 
and falls off to higher $\chi$ as $\chi^{-1/3}$.

\begin{figure}
 \includegraphics[width=\textwidth]{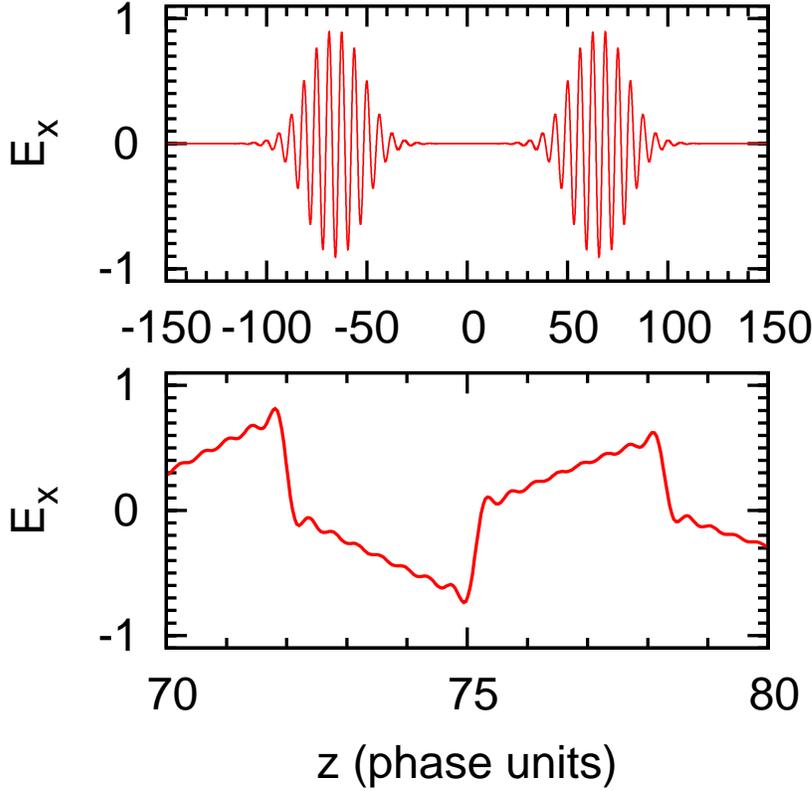}
\caption{\label{field}%
A snapshot of the 
electric field (normalized to unity at pulse centre)
used to model laser pulses of finite duration 
according to (\ref{envelopefunction}), with $L=10\pi$ 
and $\eps=10\pi/3$ (upper panel).
At $t=0$ the leading edges of the pulses meet, the snapshot
is taken at $t=-50$, corresponding to $26.5\,\textrm{fsec}$ before
interaction for a laser wavelength of $1\,\mu\textrm{m}$. 
The lower panel shows the model waveform 
used for a reflected pulse, given by (\ref{baevafield}),
with $n_{\rm max}=7$.
}
\end{figure}

\section{Trajectories}
\label{trajectories}
\subsection{Laser fields}
\label{laserfields}

Bell \& Kirk \cite{bellkirk08} considered a particle trajectory in
counter-propagating monochromatic waves, with circular polarization, such 
that the field vectors rotate together at each point in space. 
This special case is considered in detail in Section~\ref{longpulses}. 
However, such an approach is limited in three respects:
\begin{enumerate}
\item
Real laser pulses are of finite duration, the electrons they accelerate
begin and end their trajectories outside of the pulse trains. 
\item 
Circular polarization is a special case, that may not 
be the most favourable for pair production.
\item
Monochromatic waves are an idealization. In particular, if one of the 
counter-propagating waves is produced by reflection from a solid surface, 
it will contain many high-order harmonics \cite{baevaetal06}.
\end{enumerate}

We lift these limitations by considering a range of laser waveforms
as follows:
\begin{enumerate}
\item
To model pulses of finite duration, we multiply the monochromatic wave
by an envelope function $f(\phi)$ that contains two 
parameters: the duration or length $L$ of the pulses (in phase units) 
and the thickness $\eps$ (also in phase units) of the pulse edges:
\eqb
f_\pm(\phi)&=& \frac{1}{4}
\left[1\mp \tanh\left(\frac{\phi}{\eps}\right)\right]
\left[1\pm \tanh\left(\frac{\phi \pm L}{\eps}\right)\right]
\label{envelopefunction}
\eqe
Here $\phi=\phi_\pm=z\mp t$ is the phase of the wave, 
and the upper (lower) sign
refers to the rightwards (leftwards) propagating pulse.
\item
In order to avoid the stable $E=0$ nodes of the circularly polarized waves,
we consider counter-propagating, linearly polarized 
waves with {\em aligned} polarization and with {\em crossed}
polarization. In aligned polarization the electric fields of the
 counter-propagating beams are parallel, and in crossed polarization
they are orthogonal.
\item
The easiest way to produce counter-propagating waves 
may be to reflect a wave from a solid target. To account for the 
complex harmonic structure
of the reflected wave \cite{baevaetal06,baeva08}
we consider a leftward propagating wave with electric field:
\eqb
\bm{E}&=&\bm{\hat{x}}\frac{2}{\pi}\sqrt{\frac{\sqrt{3}}{2}}
f\left(\phi_-\right)  
\left\lbrace
\sum_{n=0}^{n_{\rm max}}\frac{\sin\left[\left(2n+1\right)\phi_-\right]}{2n+1}
-\frac{2\cos\left[\left(2n+1\right)\phi_-\right]}{\pi(2n+1)^2}
\right\rbrace
\label{baevafield}
\eqe 
This field is illustrated in figure~\ref{field}. The Fourier
series represents the summation of top-hat and saw-tooth functions.
Terminating the expansion at $n=n_{\rm max}$ produces the high
frequency ripple seen in the figure.
\end{enumerate}

\subsection{Equations of motion}
\label{eq_of_motion}

The trajectory is treated classically: 
the electron is taken to be a point particle
moving in prescribed external electromagnetic
fields associated with the laser beam, and radiating energy 
continuously rather than in discrete jumps.
When the electron becomes relativistic, 
radiation reaction plays an important role in the dynamics, and we incorporate 
it using the Landau-Lifshitz prescription \cite{landaulifshitz75}. 
To lowest 
order in $1/\gamma$, where $\gamma$ is the Lorentz factor,  
this leads to a force $\bm{f}_{\rm rad}$ that is
anti-parallel to the particle momentum $\bm{p}$. Writing this in terms
of the unit vector in the direction of the momentum, 
$\bm{p}=p\bm{\mu}$ one has:
\eqb
\bm{f}_{\rm rad}&=& - \frac{2e^4}{3m^4c^5}
\left|F_{\mu\nu}p^\nu\right|^2 \bm{\mu}
\\
&=&-\frac{2}{3}\alphafine\frac{m^2c^3}{\hbar}\eta^2\bm{\mu}
\label{rr_force}
\eqe
It can be shown \cite{klepikov85}
that the Landau-Lifshitz prescription for the 
classical radiation 
reaction force (the right-hand side of (\ref{rr_force}) does not
contain $\hbar$, since $\alphafine\propto 1/\hbar$ and $\eta\propto\hbar$)
is the lowest-order approximation in the small parameter
$\eta\alphafine$ to the 
physically acceptable solution of the exact Lorentz-Abraham-Dirac
equation. As such it is valid up 
to field strengths/energies much
higher than those of interest 
here \cite{dipiazza08,bellkirk08}. 
The force corresponds exactly to the 
rate of transfer of momentum from the electron to the photon field
according to the classical formula (\ref{synchclassical}).
However, the concept of a 
continuous classical trajectory fails at $\eta\approx1$, where
a single synchrotron photon takes off a significant fraction of the electron
energy. Furthermore, quantum effects reduce 
the average power radiated, modifying it by the function
$g(\eta)$ defined in (\ref{defg}) and plotted in figure~\ref{synch_g}.
In this paper, we neglect the quantum fluctuations in the electron orbit, but
take account of the reduction in radiated power by multiplying the 
radiation reaction force given in (\ref{rr_force}) by $g(\eta)$.
The equations of
motion for a particle of charge $q$, mass $m$ are then:
\eqb
\beta \frac{\diff \bm{\mu}}{\diff t}&=&
\frac{q}{\gamma m c}\left[ 
\bm{E}_\bot+\beta\bm{\mu}\wedge\bm{B}\right]
\nonumber
\\
\frac{1}{\gamma}\frac{\diff\gamma}{\diff t}&=&
\left(\frac{q}{\gamma m c}\right)\beta\bm{\mu}\cdot\bm{E}-
\frac{mc^2}{\hbar}
\frac{2\alphafine \eta^2g(\eta)}{3 \gamma}
\label{eqsmotion}
\eqe
where $c\beta=c\left(\gamma^2-1\right)^{1/2}/\gamma$ 
is the particle three speed,
$\bm{E}_\bot$ is the component of $\bm{E}$ perpendicular to $\bm{\mu}$,
and the radiation reaction term is given to lowest order in $1/\gamma$.
Note that the parameter $\eta$ is determined by the component of the 
Lorentz force perpendicular to $\bm{\mu}$:
\eqb
\eta&=&\frac{\gamma}{E_{\rm crit}}
\left[\left(\bm{E}_\bot
+\beta\bm{\mu}\wedge\bm{B}\right)^2+\left(\bm{\mu}\cdot\bm{E}\right)^2/\gamma^2\right]^{1/2}
\,\approx\,\frac{\gamma\left|\bm{E}_\bot
+\bm{\mu}\wedge\bm{B}\right|}{E_{\rm crit}}
\label{etadef2}
\eqe
Given $\bm{E}$ and $\bm{B}$ as functions of position and time,
equations~(\ref{eqsmotion}) are integrated forward in time 
using a standard fourth-order Runge-Kutta algorithm.

\begin{figure}
 \includegraphics[width=\textwidth]{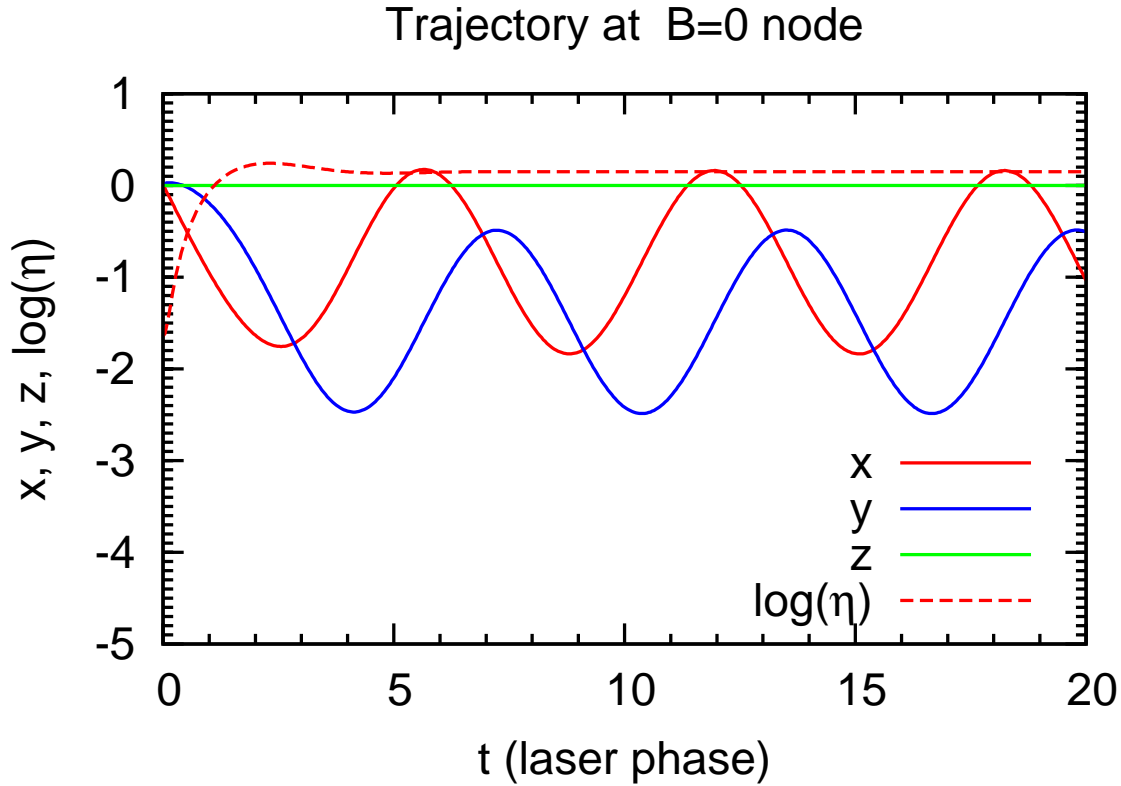}
\caption{\label{traj_circle}%
The trajectory of a particle initialized at a magnetic node in 
very long counter-propagating, circularly polarized pulses. Also plotted
is the dependence of the parameter $\eta$ on time (measured in units of 
the laser phase).
}
\end{figure}

\subsection{Pair creation}
\label{paircreation}
In addition to the electron's trajectory, we are interested also in
the number and frequency of photons it radiates and, in particular, in the
number of pairs created both by these photons as they propagate out through the
laser beams and by direct trident pair production, according to
(\ref{virtualrate}). 

The number of directly created, trident pairs is easily found
by adding equation (\ref{virtualrate}) to the set (\ref{eqsmotion}). This
describes the monotonic growth in time of the number of pairs created by
intermediate virtual photons $N_\pm(t)|_{\rm virtual}$.  

\begin{figure}
 \includegraphics[width=\textwidth]{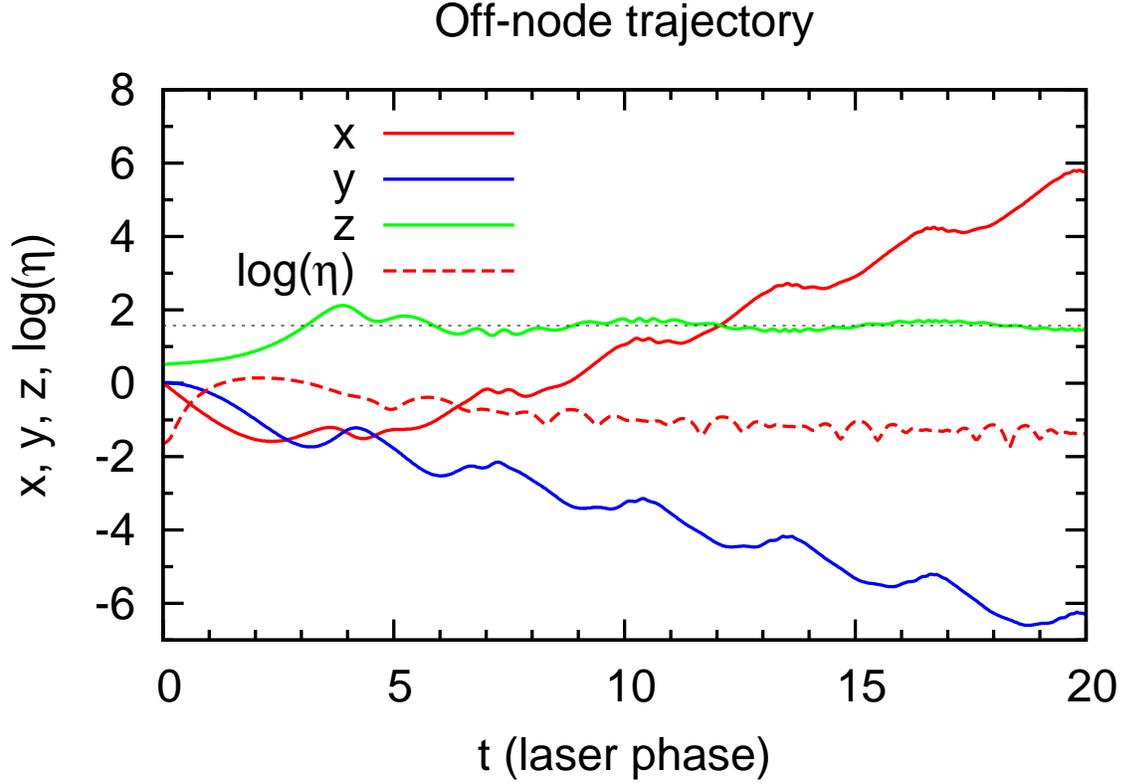}
\caption{\label{traj_off-node}
The trajectory of a particle initialized at $z=0.5$, between the 
nodes of the standing wave, in 
very long counter-propagating, circularly polarized pulses. 
The trajectory moves to a node of the electric field, located at 
$z=\pi/2$ (indicated by the dotted line).
Also plotted
is the dependence of the parameter $\eta$ on time (measured in units of 
the laser phase).
}
\end{figure}

Pairs created by 
real photons are more difficult to compute. 
Suppose an electron emits a photon at time $t_0$, position $x_0,y_0,z_0$ 
with frequency $\nu$ (all measured in the lab.\ frame). 
The total optical depth to absorption for this photon is
\eqb
\tau(\nu,t_0,x_0,y_0,z_0)
&=&\alphafine\frac{mc^2}{\hbar}\frac{mc^2}{h\nu}\int_{t_0}^{t_{\rm esc}}
\diff t \,\chi(t) T_\pm[\chi(t)]
\label{depthintegral}
\eqe
where the integration is along the photon's ray path from emission to 
escape from the system at time $t_{\rm esc}$, and $T_\pm$ is 
defined in (\ref{photonabsorption2}). 
In general, $\chi(t)$ 
depends on the photon's frequency $\nu$ and its direction, 
which are constant along the ray path, as well as on the local electromagnetic
field, which varies along this path. In analogy with (\ref{etadef2}),
$\chi$ may also be defined as
\eqb
\chi(t)&=&\frac{h\nu}{2mc^2}
\frac{\left|\bm{E}_\bot+\hat{\bm{k}}\wedge\bm{B}\right|}{E_{\rm crit}}
\label{chidef2}
\eqe
where now $\bm{E}_\bot$ refers to the 
component of the electric field perpendicular to the propagation 
direction $\hat{\bm{k}}$ of the photon. 
Therefore, assuming the photon is emitted in the direction of
motion of the electron, 
from (\ref{etadef2}) and (\ref{chidef2}) 
we can write the photon frequency in terms of its $\chi$-value at birth
and the parameters $\gamma$ and $\eta$ of the parent electron at 
that instant:
\eqb
\frac{h\nu}{2mc^2}&\approx& \frac{\chi\left(t_0\right)\gamma}{\eta}
\label{forwardpropagation}
\eqe
The value of $\chi$ at any point on the ray path is then 
easily found from the local value of the electromagnetic fields, 
since both $\nu$ and $\hat{\bm{k}}$ are constant along this path.   
The total pair production probability per electron (via real photons) 
is then given by an integral of the 
pair production rate over the frequency of the emitted photons.
The number of these pairs can also
be described by a monotonically increasing function $N_\pm(t)|_{\rm real}$
of the time $t$ at which the pair-creating photons are emitted. 
This can be evaluated by adding another equation to the set (\ref{eqsmotion}):
\eqb
\left. \frac{\diff N_\pm}{\diff t}\right|_{\rm real}&=&
\int_0^{\eta/2}\diff\chi\,
\frac{\diff N}{\diff \chi\diff t}\left[1-\textrm{exp}\left(-\tau\right)\right]
\label{realpairrate}
\eqe

\begin{figure}
 \includegraphics[width=\textwidth]{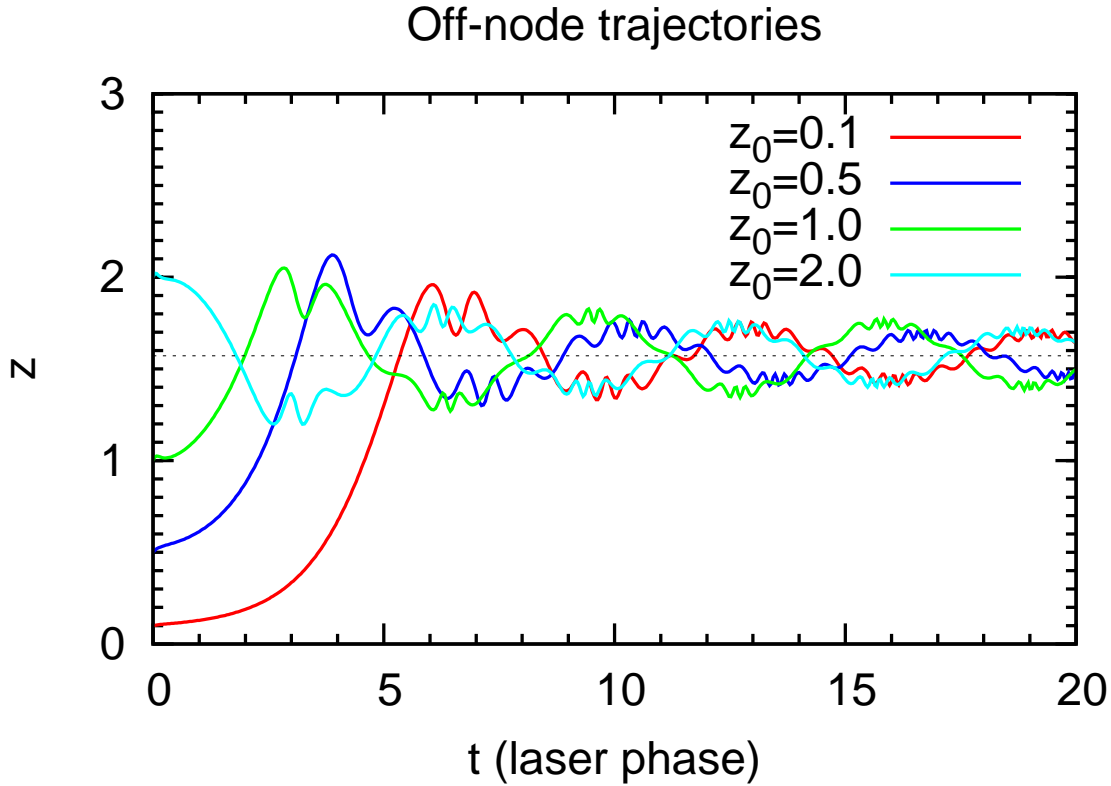}
\caption{\label{traj_off-node2}%
Trajectories of particles initialized at various positions 
in the standing wave in very long, counter-propagating, circularly
polarized pulses. 
The $B=0$ node at $z=0$ appears to be unstable, whereas
particles are attracted to the $E=0$ node at $z=\pi/2$.
}
\end{figure}

Because the conversion probability, $1-\textrm{exp}(-\tau)$,
depends on the photon's path in the strong electromagnetic fields
of the laser, a
precise treatment requires a model of the spatial 
structure of the laser beams transverse to their direction of
propagation. 
In this paper, we assume the pulses have cylindrical geometry. 
Taking the $z$-axis as the direction of propagation, the fields are
assumed to be independent of $x$ and $y$ provided $x^2+y^2\le R^2$, where
$R$ is the radius of the cylinder, which we here set equal to one 
laser wavelength: $R=2\pi$, and to vanish outside this cylinder. 
In addition, we formally 
impose an upper limit on the length 
of the cylinder $\left|z\right|\le z_{\rm max}$, although this 
is not relevant for the trajectories discussed in this paper. 
The cylindrical geometry of the pulses enters into the computations in 
two ways: particle trajectories are terminated when they leave the cylinder,
and pair creation by photons is switched off when they leave the cylinder.
We ignore diffraction effects by treating the wave as planar within the 
cylinder, but this will have little effect on the overall results. The main
motivation for imposing the cylindrical boundary
is to take approximate quantitative account of the 
escape of electrons and photons from the strong field regions
in the intersecting beams. 

\begin{figure}
\includegraphics[width=\textwidth]{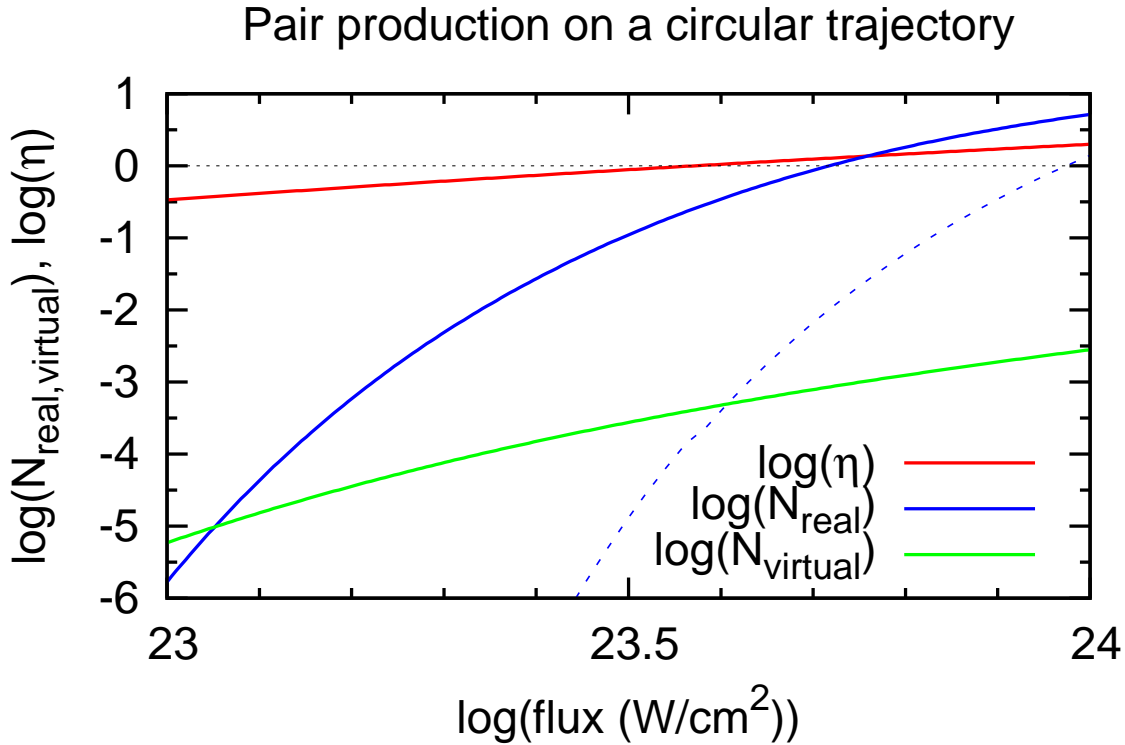}
\caption{\label{params_circle}%
The number of pairs produced during a laser period 
by an electron following a circular
trajectory at the magnetic node of the standing wave, as
a function of the intensity of (one of)
the counter-propagating, circularly polarized beams. Both the virtual
and real photon channels are shown. In addition, the 
parameter $\eta$ defined in (\ref{etadef}), 
which is constant on this trajectory, is shown.
For comparison, the number of pairs 
predicted in \cite{bellkirk08}
is shown as a dashed line.
}
\end{figure}

This procedure should yield a realistic estimate of the 
number of pairs created by a single electron and the photons it radiates. 
However, in section~\ref{longpulses}, for the purpose of comparison  
we adopt
an approximate method of 
computing the optical depth to absorption $\tau$ that is
closely similar
to the method used by Bell \& Kirk \cite{bellkirk08}.
The approximation consists of 
replacing the integral over the photon trajectory
in (\ref{depthintegral}) according to
\eqb
\int_{t_0}^\infty
\diff t \,\chi T_\pm(\chi)
&\rightarrow&
\frac{L_\pm}{c}
\left<\chi\right> T_\pm\left(\left<\chi\right>\right)
\label{approxdepthintegral}
\eqe
The two parameters $\left<\chi\right>$
and $L_\pm$ can be interpreted as the effective value of $\chi$ 
averaged along the escaping ray path and the
effective length of this path (assumed to be 
the same for each photon independent of its frequency or the position
at which it is created). 

As it propagates through the laser beams, 
the photon encounters electromagnetic
fields that vary in both magnitude and direction. 
If, for example, the photon is emitted
almost parallel to the local electric field, the initial
value of $\chi$ is much lower than the average value it experiences 
on its escape path. To allow for this, we choose the average value 
$\left<\chi\right>$ so that it corresponds to a photon propagating 
perpendicular to an electric field whose strength equals 
the amplitude of the oscillating field at the point of emission.
Formally, this involves summing the envelope functions 
($f_+$ and $f_-$) introduced 
in (\ref{envelopefunction}):
\eqb
\lefteqn{%
\left<\chi\right>\left(\nu,t_0,x_0,y_0,z_0\right)\,=\,}&&
\nonumber\\
&&
\frac{h\nu}{2mc^2}b
\left[f_+(k_{\rm laser}z_0-\omega_{\rm laser} t_0) + 
f_-(k_{\rm laser}z_0+\omega_{\rm laser} t_0)\right]
\eqe
so that, using (\ref{depthintegral}) and (\ref{approxdepthintegral}), 
the approximate optical depth becomes
\eqb
\lefteqn{%
\left<\tau\right>\left(\nu,t_0,x_0,y_0,z_0\right)\,=\,}&&
\nonumber\\
&&\frac{L_\pm}{2}\alphafine\strength T_\pm\left(\left<\chi\right>\right)
\left[f_+(k_{\rm laser}z_0-\omega_{\rm laser} t_0) + 
f_-(k_{\rm laser}z_0+\omega_{\rm laser} t_0)\right]
\label{avdepth}
\eqe
This quantity then replaces $\tau$ in (\ref{realpairrate}). 
In \cite{bellkirk08}, the effective path length was
estimated to be one laser wavelength, so that 
we set $L_\pm= 2\pi c/\omega_{\rm laser}$.

\begin{figure}
\includegraphics[width=\textwidth]{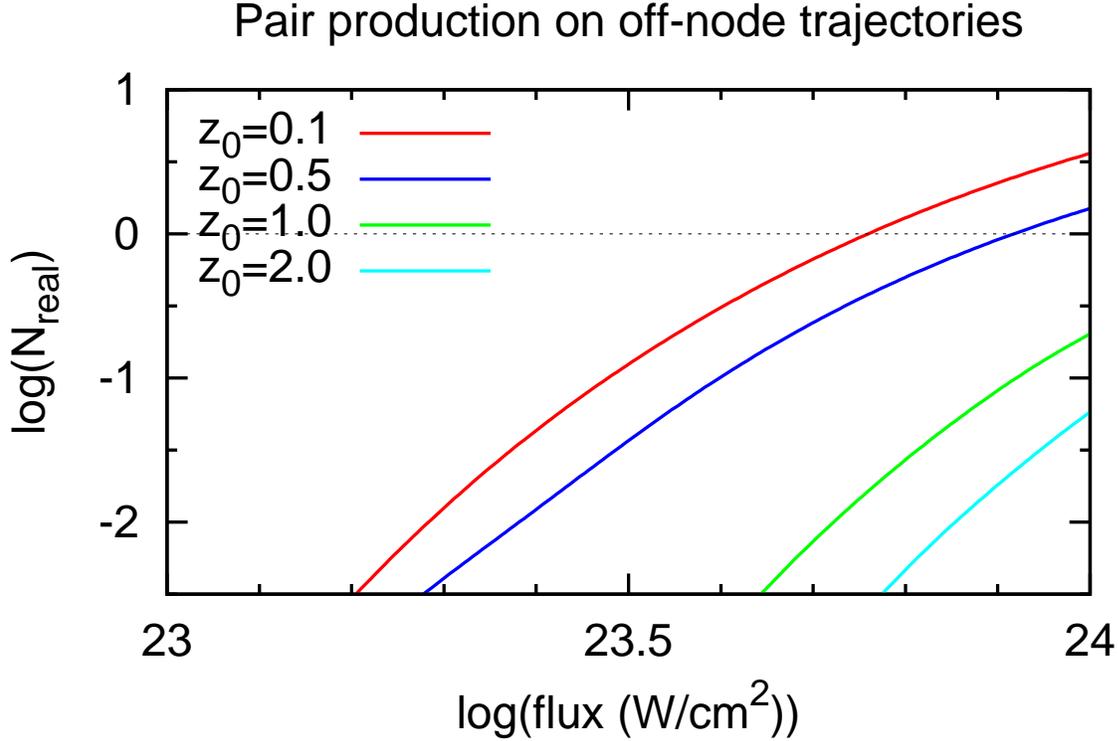}
\caption{\label{params_off-node}%
The number of pairs produced at $t=20$ 
(by which time this quantity has saturated)
by an electron initialized at $t=0$ between the nodes of 
the standing wave, as
a function of the intensity of (one of)
the counter-propagating, circularly polarized 
beams. Only the real photon channel is
shown. Initial conditions of the trajectories are as in 
figure~\ref{traj_off-node2}
}
\end{figure}

\section{Results}
\label{results}
Our results are divided into two parts. In the first, we consider
counter-propagating, circularly polarized pulses of infinite duration. 
This enables us to compare with the results of \cite{bellkirk08}. Pair 
production in this case is treated using the approximation given in 
(\ref{approxdepthintegral}). In the second part, 
section~\ref{shortpulses}, we consider a more realistic model, 
with pulses of finite duration. Pair 
production in this section is computed by integrating along the escape path
of each photon at each timestep, and the laser fields are assumed to be 
confined to a cylindrical region. 

\subsection{Circularly polarized, long duration pulses}
\label{longpulses}

Bell \& Kirk \cite{bellkirk08} analyzed a particularly simple
trajectory: assuming two long pulses $L\rightarrow\infty$, they found
that at the magnetic node ($B=0$), the particle settles into a
circular orbit. This is illustrated in figure~\ref{traj_circle}.  The
electron starts at $t=0$ at the origin, which is a node of the $B$
field, with velocity directed along the positive $y$ axis, and Lorentz
factor $\gamma=10$. The intensity of a single laser pulse in this and
the following examples is $6\times10^{23}\,\wsqcm$. Within less than 
one hundredth of a laser period the particle turns to move in
the negative $x$-direction, and, after a transient
lasting less than one half of a laser period, the trajectory 
relaxes to a 
circle in the $x$-$y$ plane, with constant Lorentz factor
$\gamma=760$.

Initializing an electron at a magnetic node is convenient for analytical
calculations. However, it selects an electron trajectory that 
is particularly efficient at pair production. An electron initialized at rest 
at a node of the electric field, for example, remains at rest indefinitely.
If a starting point in between the electric and magnetic nodes is chosen, 
the electron tends to drift towards the electric node, and lose energy in
the process. This is illustrated in figure~\ref{traj_off-node}, which shows
the trajectory of an electron initialized at $z=0.5$ (in units of 
$\lambda_{\rm laser}/2\pi$), 
with the same velocity as in figure~\ref{traj_circle}.
As well as a drift perpendicular to the laser beams towards 
positive $x$ and negative $y$,
this trajectory starts to oscillate around the $E=0$ node at $z=\pi/2$,
and slowly converges onto it. After a sharp rise associated with
the rapid initial acceleration experienced by the electron, the 
parameter $\eta$, that controls pair production, 
decreases steadily with time.

\begin{figure}
\includegraphics[width=\textwidth]{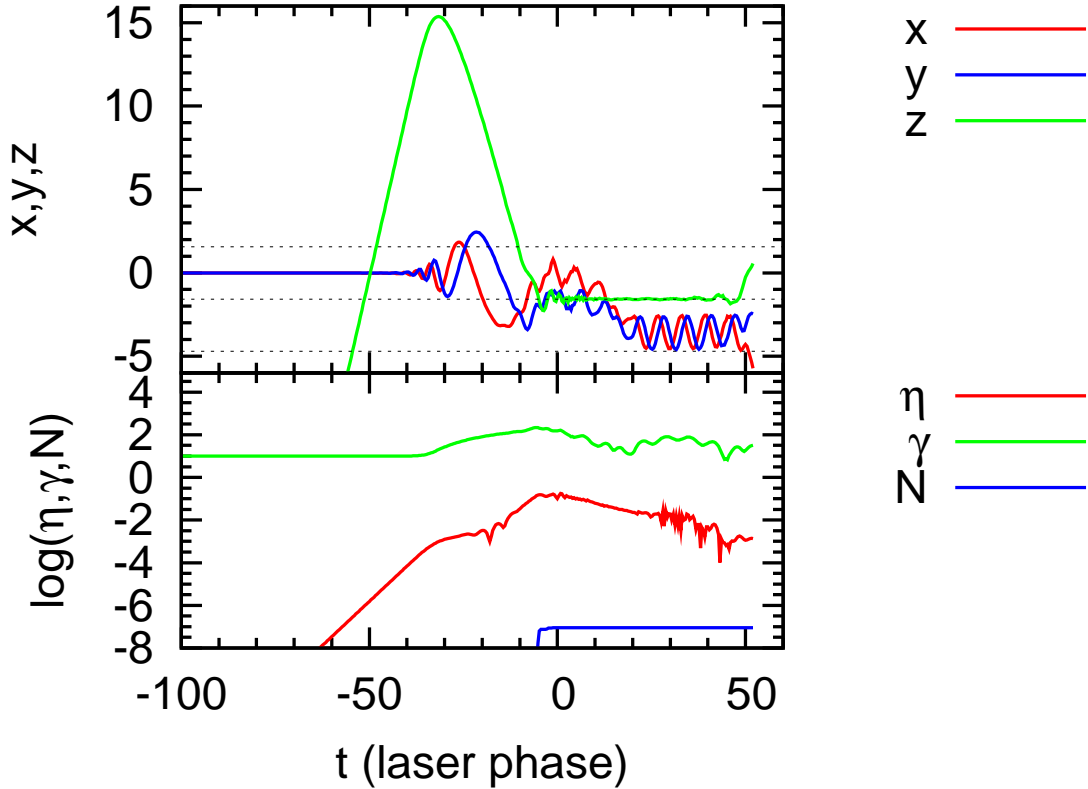}
\caption{\label{circularpol_traj}%
Sample trajectory in circularly 
polarized pulses of finite length, propagating along the $z$-axis.
The upper panel shows position, the lower panel the Lorentz factor, 
the number $N$ of 
pairs produced per primary electron and the parameter $\eta$
defined in (\ref{etadef}). 
When the pulses overlap the 
electric field vanishes at $z=(2n+1)\pi/2$, shown by dashed lines
for $n=-2$, $-1$ and $0$.
}
\end{figure}

Convergence on the $E=0$ node is a general property of the trajectories, as
shown in figure~\ref{traj_off-node2}. 
These examples demonstrate that the majority
of the trajectories move swiftly to the vicinity of the $E=0$ mode,
arriving in a zone around it of dimension less than one tenth of a wavelength
within a single laser period. The subsequent motion is complex. As the 
electrons lose energy, their natural time and length scales 
decrease. After a long period close to 
such a node, this behaviour can place strong demands on the numerical 
integration method. 

The number of pairs produced by a single electron following the 
trajectory illustrated in figure~\ref{traj_circle} 
for a range of laser intensities
is shown in figure~\ref{params_circle}. 
The initial conditions were chosen to place the particle on the 
circular path analyzed in \cite{bellkirk08}, 
thus avoiding short-lived transients. On such a 
trajectory, the parameter $\eta$ is constant. The number 
of pairs increases linearly with time, 
but, for comparison with \cite{bellkirk08}, 
we select the value after one laser-period
($t=0$ to $t=2\pi$).
This figure was computed using the 
approximate description of pair production described in 
section~\ref{paircreation},
with $L_\pm=2\pi c/\omega_{\rm laser}$. 
It is equivalent to the method used 
by Bell \& Kirk \cite{bellkirk08} in their figure~(1), in which,
in addition, 
a monochromatic approximation to the synchrotron emission was adopted. As
expected, allowing for the fact that even low energy electrons occasionally
emit photons capable of pair producing leads to an increase in the number
of pairs produced at low laser intensities. As a result, the pairs
produced via virtual photons are overwhelmed by those produced by real photons,
except at 
$I<10^{23}\,\wsqcm$, where the production rate is in any case very low.
An additional factor contributing to the higher rate of pair 
production found in these calculations when compared to \cite{bellkirk08} is
the inclusion of quantum corrections to the radiation reaction force 
via the factor $g(\eta)$ in (\ref{eqsmotion}). This leads to larger 
values of both $\gamma$ and $\eta$ for given laser intensity.

If the electron is initialized in between the nodes, pair production is 
reduced. This is shown in figure~\ref{params_off-node}. In contrast with
the circular trajectory at the $B=0$ node, the number of pairs produced when
the particle is initialized off-node does not increase linearly with time, but
saturates after a few laser periods. Consequently, figure~\ref{params_off-node}
shows the {\em time-asymptotic} number of pairs, for the four 
initial conditions shown in figure~\ref{traj_off-node2}.  

The curve for $z_0=2.0$ demonstrates that, for circular polarization,
electrons initialized close to the $E=0$ node move towards it rapidly
without emitting as many pairs as those electrons initialized close to
the $B=0$ node.  Electrons with an unfavourable initial position do not
contribute substantially to pair production.  However,
this negative effect is offset by the positive 
modifications illustrated in figure~\ref{params_circle},
The general conclusion of 
\cite{bellkirk08}
that an ensemble of $N$ electrons produces approximately $N$ 
secondary pairs when
the intensity of circularly polarized counter-propagating laser beams
into which they are injected
reaches $10^{24}\,\wsqcm$ remains unchanged.

\begin{figure}
\includegraphics[width=\textwidth]{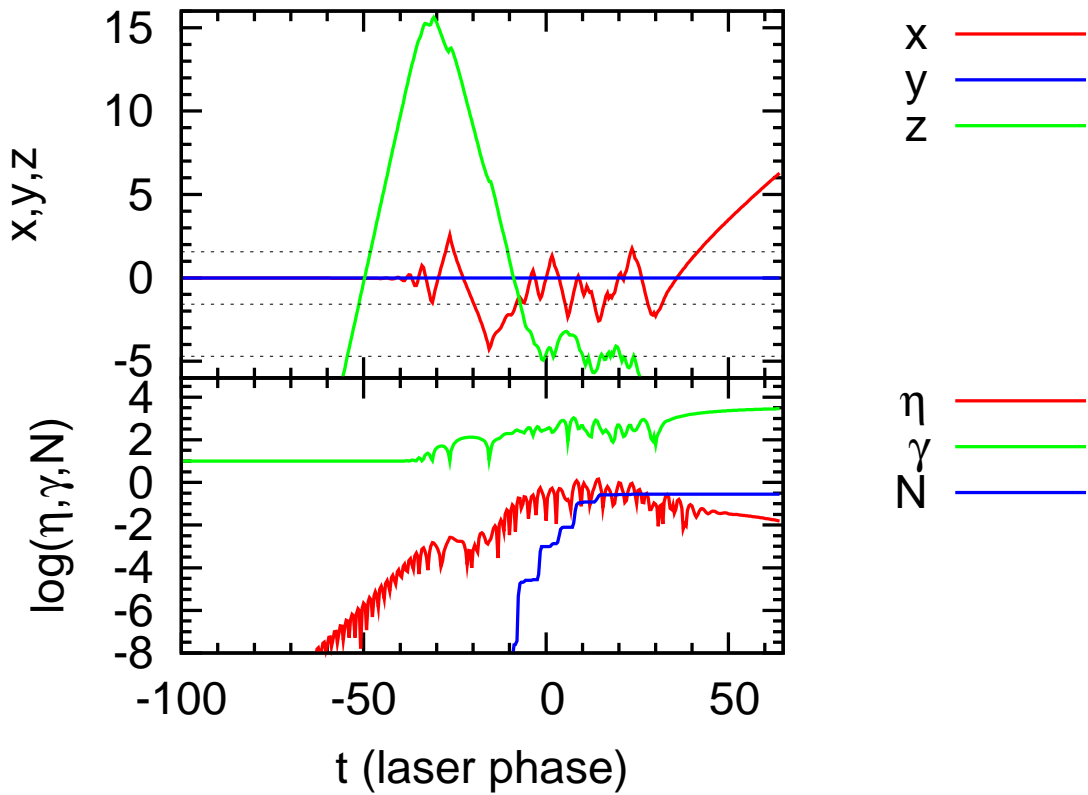}
\caption{\label{linearpol_traj}%
Sample trajectory in 
finite-length pulses with aligned, linear polarization. 
}
\end{figure}

\subsection{Finite duration pulses in cylindrical geometry}
\label{shortpulses}

The results of the previous section show that the position of a particle 
within a circularly polarized, 
long pulse train is important in determining the number of pairs 
it produces. In reality, however, pulse lengths are limited to a few laser 
wavelengths. As the pulses 
approach each other they pick up electrons and carry them 
forward on the front edge of the pulse until interaction starts. 
During interaction, the electron trajectory remains
for an extended period at approximately constant $z$. 

In the case
of circular polarization, the particle stays close to a node of 
the electric field, but in the case of linear polarization, 
it samples a much larger region. 
In general terms, this is due to the characteristic 
figure-of-eight
oscillation of an electron in a linearly polarized wave during which
the electron oscillates in the direction parallel to beam propagation
as well as in the perpendicular direction. In contrast, in a
circularly polarized wave the electron can move smoothly, without
parallel oscillation, towards the attracting $E=0$ node.
This behaviour is illustrated in 
figures~\ref{circularpol_traj} and \ref{linearpol_traj},
in which the laser intensities and laser pulse shapes are the same
but the polarizations differ.  
In each case,
pulses of peak intensity $I_{24}=0.6$ 
with length $L=10\pi$ and thickness of the pulse edge
$\eps=10\pi/3$ were chosen. The trajectory was initialized at 
$t=-100$, when the pulses are well separated, at the position
$x=y=0$, $z=-50$, moving in the positive $z$ direction with 
Lorentz factor $10$. The particle coasts at constant
$\gamma$ until $t\approx-30$, when it encounters the leading edge of the 
leftward propagating wave. This wave picks it up, reversing its motion along
$z$ and accelerating it until the pulses begin to interact at $t\approx0$.
In the circularly polarized case, the particle immediately drops onto 
the $E=0$ node at $z=-\pi/2$ (shown as a dashed line), the Lorentz factor falls
and few pairs are created ($\sim10^{-7}$ per primary particle). However, in the 
linearly polarized case, the particle performs substantial excursions in 
$z$, settling only briefly onto the $E=0$ node at $z=-3\pi/2$. 
During this time, the Lorentz factor fluctuates rapidly, with only 
a slight decrease in its average value. 
As a result, a relatively
large number ($0.3$ per primary particle) of pairs are created. The 
trajectories terminate when the particle exits the cylinder containing
the laser pulses at roughly $t=50$ for circular polarization and $t=65$ for 
linear polarization.

During the period when the pulses overlap, the electron trajectory 
is very complex, especially in the case of aligned linear polarisation. 
This makes it quite sensitive to the accuracy demanded of the integration 
algorithm, in the sense that a more stringent accuracy 
requirement leads to a trajectory that remains converged to larger times. 
For the trajectory shown in 
figure~\ref{linearpol_traj}
we find that an error per step of less than
$\delta =10^{-4}$ in position and in the angles (in radians) 
used to describe the unit vector $\bm{\mu}$ yields a converged trajectory 
up to $t=20$, whereas $\delta=10^{-6}$ extends this to $t=35$. (The example 
presented in figure~\ref{linearpol_traj} was computed with $\delta=10^{-8}$.)
However, in both cases pair creation
ceases well before accuracy is lost. We conclude that 
counter-propagating, 
circularly polarized beams are 
intrinsically inefficient at pair creation, at least when the fields rotate 
in the same direction, and do not consider them further.   
\begin{figure}
\includegraphics[width=\textwidth]{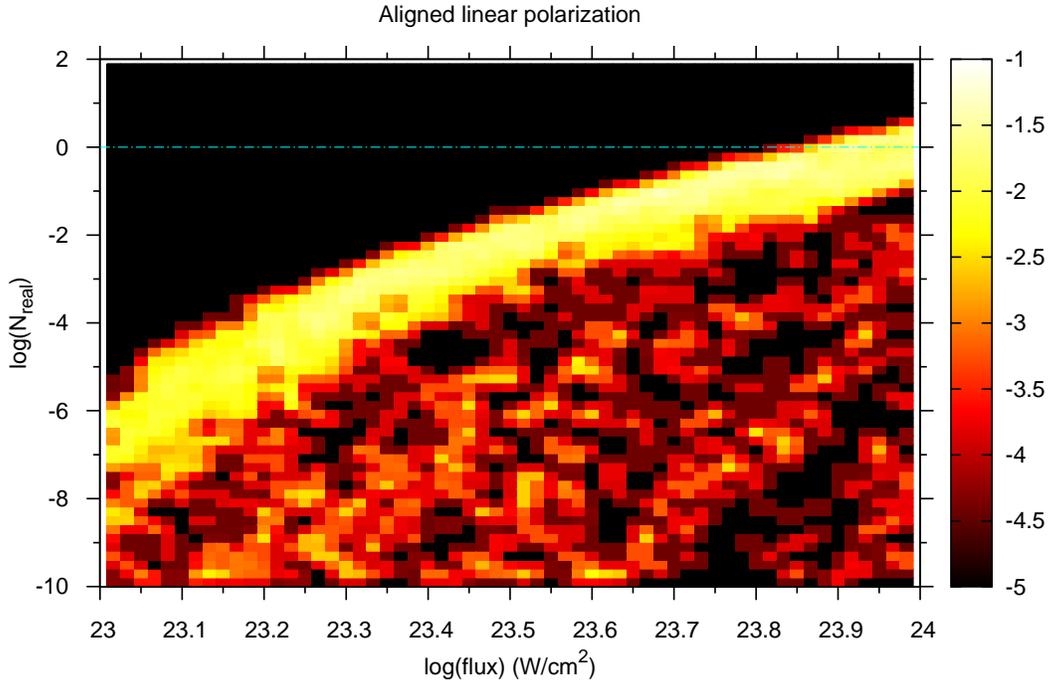}
\caption{\label{monte_aligned}%
Pair production rates for a range of initial particle positions
and for aligned linear polarization. 
Each pixel corresponds to a small range of laser intensity 
(\lq\lq flux\rq\rq, shown on the $x$-axis) and
of the number $N_{\rm real}$ of pairs produced via the channel that involves 
an intermediate real photon (shown on the $y$-axis). 
The colour coding of 
a pixel indicates the
logarithm (to base 10) of the probability that, 
given a flux in this interval, 
an electron with randomly chosen 
initial conditions (see text) will produce  
the corresponding number $N_{\rm real}$ 
of secondary pairs.
Roughly $80\%$ of trajectories produce no pairs and so are not shown.
At a laser flux of $10^{24}\,\wsqcm$, 
about $10\%$ of the particles 
reach the interaction region and the majority of these 
produce on average roughly one secondary pair.
}
\end{figure}

\begin{figure}
\includegraphics[width=\textwidth]{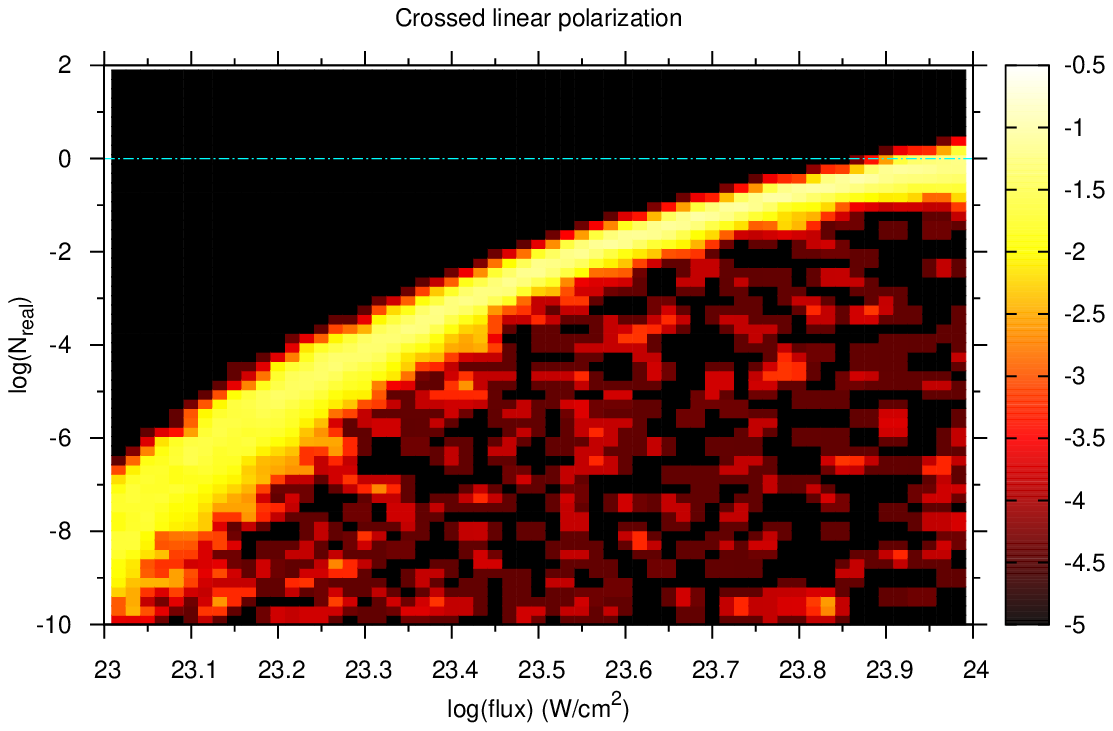}
\caption{\label{monte_crossed}%
Pair production rates for a range of initial particle positions,
as in figure~\ref{monte_aligned}, but for crossed linear polarization
}
\end{figure}

\begin{figure}
\includegraphics[width=\textwidth]{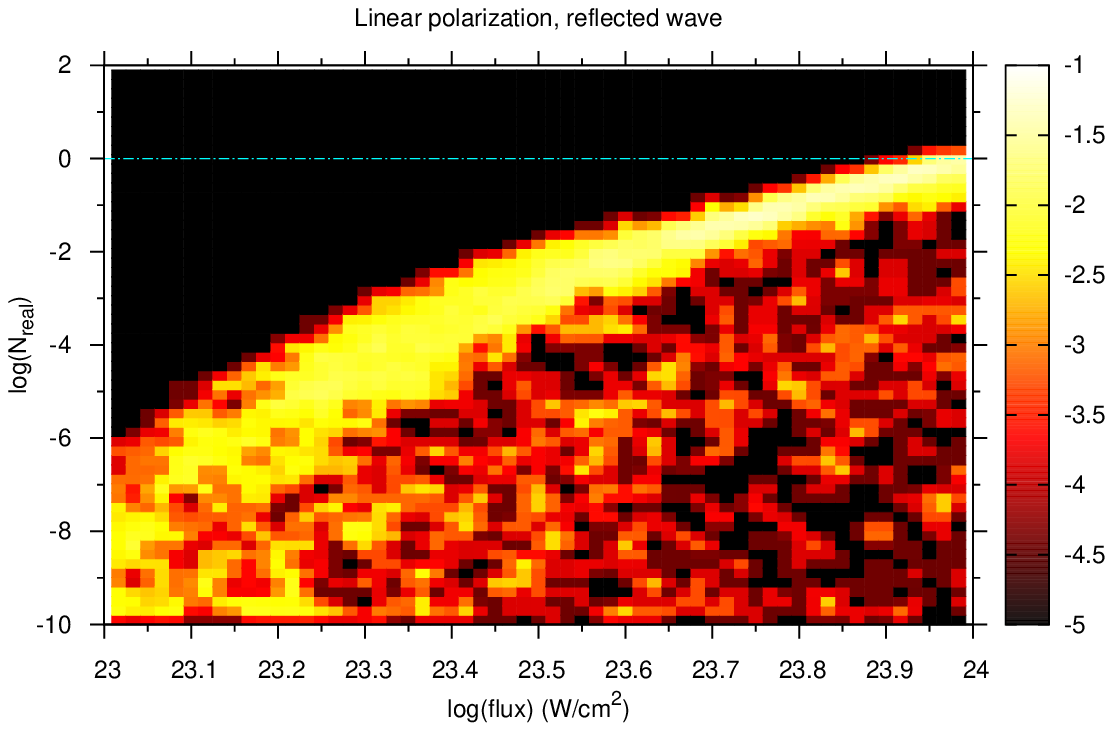}
\caption{\label{monte_baeva}%
Pair production rates for a range of initial particle positions
as in figure~\ref{monte_aligned}, but 
for aligned linear polarization with 
the leftward propagating wave modified to represent
a reflected wave (with $n_{\rm max}=7$, see (\ref{baevafield})).
}
\end{figure}

Trajectories in linearly polarized 
pulses of finite length, contained in the cylindrical
volume described above are sensitive to the point at which they are
first picked up by a pulse, and to the time they subsequently spend at
its leading edge, before they encounter the counter-propagating
pulse. To take account of this, we present, in
figures~\ref{monte_aligned}, \ref{monte_crossed} and \ref{monte_baeva},
results obtained in pulses with $L=10\pi$ and $\eps=10\pi/3$
by varying the initial $x$, $y$ and $z$ positions of
the particle, whilst keeping the other initial conditions fixed. 
In each case, $3\times10^4$ trajectories were computed, using
randomly selected values of $x$ and $y$ uniformly distributed over the
cross-section of the cylinder, random values of $z$ distributed
uniformly over the range $-50$ to $0$, and random values of the pulse
intensity, distributed uniformly in logarithm over the range
$23<\log_{10}\left(I_{24}\right)<24$. 
The initial Lorentz factor is set to $10^{0.001}$, corresponding to an 
electron energy of about $1\,$keV, and the initial velocity is in the 
positive $z$-direction.
In each of the three cases,
roughly $80\,\%$ of the trajectories exit the cylinder before reaching
the region where the laser beams interact, and so fail to produce
pairs. The remaining trajectories were accumulated in 60~bins in
intensity and 60~bins in pair yield, and are displayed as colour
contour plots.

The aim in figures \ref{monte_aligned}, \ref{monte_crossed} and
\ref{monte_baeva} is to test the sensitivity of pair production to
polarization and pulse shape.  In figure~\ref{monte_crossed}, \lq\lq
crossed polarization\rq\rq, the electric field of one beam is in the
$x$-direction whereas the field of the other beam is in the
$y$-direction. In figures~\ref{monte_aligned} and \ref{monte_baeva}
the electric field of each pulse is in the $x$-direction. 
Figure~\ref{monte_aligned} and \ref{monte_crossed}
present results obtained using 
sinusoidal waves 
(modulated by the envelope function), 
whereas in figure~\ref{monte_baeva} high harmonics are included in
one of the waves, as illustrated in figure (\ref{field}).

The possible advantage of the crossed configuration is that
 only at isolated points and times is the total electric field zero.
 When considering circular polarization in section \ref{longpulses},
 it was found that electrons congregated at the $E=0$ node and pair
 production was dramatically reduced.  By eliminating the $E=0$ node
 this possibility is removed, which suggests that pair production
 might be increased.  However, comparison of
 figures~\ref{monte_aligned} and \ref{monte_crossed} shows that \lq\lq
 aligned polarization\rq\rq\ with both electric fields oscillating in
 the same direction performs nearly as well as 
crossed polarization --– possibly because the maximum electric field
 is larger by $\sqrt{2}$.  In all three cases, 
(figures~\ref{monte_aligned}, \ref{monte_crossed} and \ref{monte_baeva}) those
 trajectories that produce fewer than $10^{-10}$ pairs per primary
 electron are omitted.  The remaining trajectories, show a large range
 in the number of pairs created, particuarly for aligned polarization at 
low laser intensity.  However, most of the electrons sit
 on a band stretching from $N_{\rm real}\approx10^{-8}$ at an
 intensity of $10^{23}\,\wsqcm$ to $N_{\rm real}\approx1$ at
 $10^{24}\,\wsqcm$.  This implies that pair production is effective
 for those particles that reach the interaction region, although only
 a minority of electrons do so if, as we assume here, they are
 injected uniformly over the beam cross-section just before the pulses
 meet.

In figure~\ref{monte_baeva}, the hypothesis is tested that pair
production can be enhanced if the waveform contains high-frequency
harmonics.  At first sight, this is not implausible, because of the 
following argument: As discussed in
\cite{bellkirk08}, radiation losses prevent 
electrons reaching the Lorentz factor 
$\gamma_{\rm max}=eA_{\rm max}/mc^2$, 
where $A_{\rm max}$ is the maximum value of
the electromagnetic vector potential. They do this by
aligning the particle velocity and the accelerating force exerted by the
field, and, therefore, not only reduce the Lorentz factor 
below $eA_{\rm max}/mc^2$, but also reduce the perpendicular component of the
field and the parameter $\eta$ that controls both pair
production and radiation losses. Prolific pair production would be much 
easier to achieve if radiation losses could be avoided 
whilst the electron is being accelerated.
This suggests that if
one of the counter-propagating waves contained abrupt changes in the 
electric or magnetic field it might be possible for the other wave to
accelerate the electron in the direction of laser propagation to a
high Lorentz factor but relatively small $\eta$. Subsequently,
an abrupt change in the field direction caused by the 
first wave
could lead to a large value of $\eta$ before 
radiation losses set in.  

Such a situation might arise naturally when
a sinusoidal wave is incident on a solid target since, as shown by
\cite{baevaetal06}, the reflected wave contains harmonics and abrupt
changes in the electric field.  According to PIC simulations~\cite{baeva08}, 
at an intensity of around $10^{21}\,\wsqcm$, the reflected
waveform is generically similar to that in figure~\ref{field}.  
The results
presented in figure~\ref{monte_baeva} are computed using this waveform
for one of the waves while keeping the total intensity
unchanged. However, at least in this example, 
our hypothesis is not confirmed: 
the presence of
harmonics appears to make little difference and electrons produce
roughly the same numbers of pairs as in sinusoidal waves 
with aligned linear polarization.

We conclude from figures~\ref{monte_aligned}, \ref{monte_crossed} and
\ref{monte_baeva}, that pair production
is relatively insensitive to the orientation of the planes of 
linear polarization and to the waveform.  While these results
provide no indication that we can further increase pair production by
a careful choice of waveform, they reinforce our
general conclusion that strong pair production takes place whenever
counter-propagating or reflected beams reach intensities around
$10^{24}\,\wsqcm$.

\section{Conclusions}
\label{conclusions}

In the above, we examine the possibility that 
an electron moving in counter-propagating laser
beams of intensity $\sim10^{24}\,\wsqcm$ could produce a substantial
number of electron-positron pairs solely as a result of interaction
with the laser fields. This is done using 
an accurate treatment of the physical processes,
and embedding these in a numerical analysis of 
classical trajectories in 
realistic field configurations.

In the case of the photon production rate, we lift the monochromatic
approximation used in \cite{bellkirk08} and take full account of
quantum effects in the weak-field, quasi-stationary approximation
(that is well-justified in the regime of interest). The result is
shown in figure~\ref{params_circle}, where we 
examine a circular trajectory at the
$B=0$ node of counter-propagating, monochromatic, circularly polarized
beams and compare our results with \cite{bellkirk08}.  Because even
relatively low energy electrons produce a few hard photons that are
subsequently capable of pair production, pairs start to appear at even
lower intensities than previously predicted. The process of
direct production via virtual intermediate photons, which was
predicted to dominate over production via real photons for intensities
less than $3\times10^{23}\,\wsqcm$ is swamped by this effect for laser
intensities exceeding $10^{23}\,\wsqcm$. At lower intensities, the
virtual photon channel still dominates, but the overall rate of pair
production is very small.  At higher intensities, where the parameter
$\eta$ is of the order of unity, classical synchrotron theory predicts
the emission of photons whose energy exceeds that of the radiating
electron. When this is corrected, (see figures~\ref{synchrotronfig} and
\ref{synch_g}), not only is the rate of emission of hard photons
reduced, but the radiation reaction force acting on the electron is
diminished.  Radiation
reaction significantly inhibits pair production on a circular
trajectory, so that these two corrections have opposite effects on the
pair production rate. Figure~\ref{params_circle} shows that the
overall effect is a slight enhancement of the pair production rate ---
we find the threshold above which an electron on this circular
trajectory produces more than one pair per laser period is
crossed at about $5\times10^{23}\,\wsqcm$, compared to
$10^{24}\,\wsqcm$ found in \cite{bellkirk08}.

To take account of more realistic trajectories, we consider pulses of
finite length and finite cross-section, with both linear and circular
polarization, and inject electrons at a complete range of phases and
positions. In addition to sinusoidal waveforms, we examine a laser
pulse with harmonics, such as might be generated by reflection by an
overdense plasma.  The electron-laser interaction at the $B=0$ node
considered in \cite{bellkirk08} is a special case, not least because
the trajectory is unstable and the electrons migrate towards the $E=0$
node if the laser beams are circularly polarized. Nevertheless, under
a wide range of laser pulse-shapes and polarizations with injection at
random phases, we find that the conclusion is essentially
unchanged: as the intensity of counter-propagating beams approaches
$10^{24}\,\wsqcm$, the number of secondary pairs produced becomes
roughly equal to the number of primary electrons interacting with the
beams. Since each electron and positron so produced interacts further
with the laser beams to produce further pairs, an avalanche of pair
production is possible.

Further work is needed, particularly to remove the assumption that
electron radiation losses are continuous, and to follow the
interaction of primary and secondary particles with the complicated
field structures produced by laser-solid interactions. However, our
results strongly suggest that it may be possible to produce
substantial pair plasmas with the next generation of 
extremely high-power lasers.

\ack
We thank Teodora Baeva, Karen Hatsagortsyan, Carsten M\"uller
and Antonio di~Piazza for helpful discussions. 
JK thanks Katherine Blundell and the 
St John's College Research Centre, University of Oxford
for their hospitality.


\appendix
\section{Synchrotron emissivity}
\label{appendix_synchrotron}
The function $F(\eta,\chi)$ is given in equations~(2.5a\dots f) in \cite{erber66}:
\eqb
F(\eta,\chi)&=&\frac{8\chi^2}{3\sqrt{3}\pi\eta^4}\sum_{i=1}^3 F_i(2\chi/\eta)\,J_i(y)
\eqe
where 
\eqb
y&=&\frac{2\chi}{\left[3\eta\left(\eta-2\chi\right)\right]}
\label{yquantumdef}
\eqe
is the quantum equivalent of the classical parameter (photon frequency/characteristic synchrotron frequency).
The $F_i$ are all positive, provided that the electron energy exceeds
that of the photon, $\eta>2\chi$:
\eqb
F_1(x)&=&1+\left(1-x\right)^{-2}
\nonumber\\
F_2(x)&=&2\left(1-x\right)^{-1}
\nonumber\\
F_3(x)&=&x^2\left(1-x\right)^{-2}
\eqe
\begin{figure}
\includegraphics[width=\textwidth]{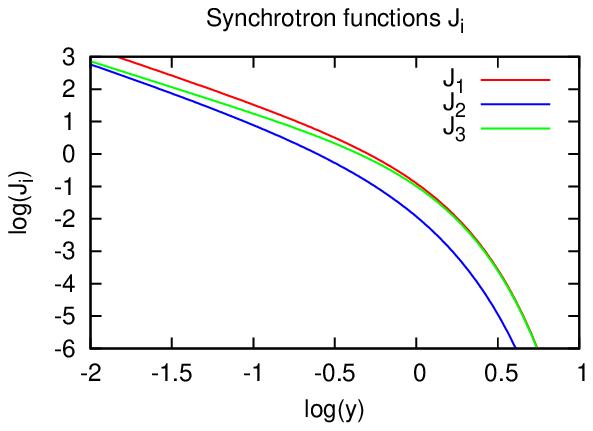}
\caption{\label{jfunctions}%
The synchrotron functions $J_i$ defined in (\ref{jfunctionsdef})
}
\end{figure}

The $J_i$ are also positive:
\eqb
J_1(y)&=&\frac{1}{3y^2}\int_y^\infty\,\diff u\,\frac{u}{\sqrt{\left(u/y\right)^{2/3}-1}}K_{2/3}^2\left(u\right)
\nonumber\\
J_2(y)&=&\frac{1}{3y}\int_y^\infty\,\diff u\,\left(\frac{u}{y}\right)^{1/3}\sqrt{\left(u/y\right)^{2/3}-1}K_{1/3}^2\left(u\right)
\nonumber\\
J_3(y)&=&\frac{1}{3y^2}\int_y^\infty\,\diff u\,\frac{u}{\sqrt{\left(u/y\right)^{2/3}-1}}K_{1/3}^2\left(u\right)
\label{jfunctionsdef}
\eqe
For $y\ll1$ they are well approximated by 
\eqb
J_1(y)&\approx&\frac{1}{3y^{5/3}}\int_0^\infty\,\diff u\, u^{2/3}K_{2/3}^2(u)
\nonumber\\
&\approx&0.921 y^{-5/3}
\\
J_2(y)&\approx&\frac{1}{3y^{5/3}}\int_0^\infty\,\diff u \,u^{2/3}K_{1/3}^2(u)
\nonumber\\
&\approx&0.307 y^{-5/3}
\\
J_3(y)&\approx&J_2(y)
\eqe
These functions are plotted in figure~\ref{jfunctions}

\begin{figure}
\includegraphics[width=\textwidth]{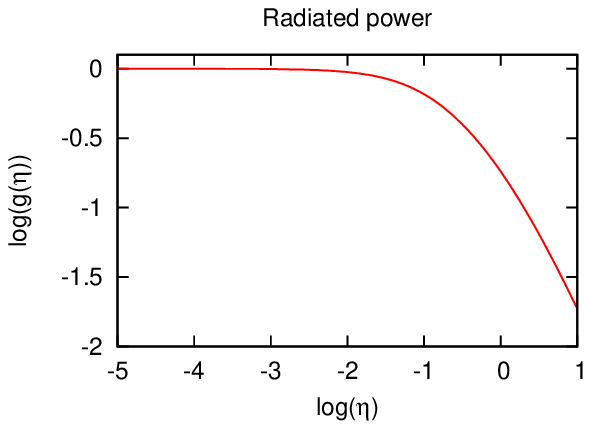}
\caption{\label{synch_g}%
The function $g(\eta)$ defined in (\ref{defg})
that describes the quantum corrections to the 
total power in synchrotron radiation
}
\end{figure}

The function $g(\eta)$ defined in (\ref{defg}) that governs the
total radiated energy can be written \cite{sokolovternov68}:
\eqb
g(\eta)&=&\frac{9\sqrt{3}}{8\pi}
\int_0^\infty\,\diff y\, \left[
\frac{2y^2K_{5/3}(y)}{\left(2+3\eta y\right)^2}+
\frac{36\eta^2 y^3K_{2/3}(y)}{\left(2+3\eta y\right)^4}
\right]
\eqe
This function is plotted in figure~\ref{synch_g}.

A closely related alternative approximation 
to the synchrotron function is given by 
\cite{akhiezeretal94}:
\eqb
F(\eta,\chi)&=&\frac{4\chi^2}{\eta^2}yK_{2/3}(y)+
\left(1-\frac{2\chi}{\eta}\right)y\int_y^\infty\,\diff t\,K_{5/3}(t)
\eqe
where $y$ is defined in (\ref{yquantumdef}).

\section{Pair-production via virtual photons}
\label{virtualappendix}
The function $\hat{\Omega}(\eta)$ can be evaluated starting from the expression
given in~\cite{erber66} (equation~A39):
\eqb
\Omega''(x)&=&4\pi\int_0^\infty\,\diff u\,K_2(4xu)K_{2/3}(1/u)
\eqe
{\em Mathematica\/} gives an explicit expression for this integral in terms 
of the Meijer~G~function, and integrates it twice 
(using the boundary conditions $\Omega'(\infty)=\Omega(\infty)=0$) to obtain
\eqb
\hat{\Omega}(\eta)&=&\frac{\pi}{16}G^{6,0}_{2,6}\left(
\frac{16}{9\eta^2}\!\!\!\!\!\!\!\!\!\!\!\!\!\!\!\!
\begin{array}{ccc}
&&\\
&&\\
&&
\end{array}
\right|\left.
\begin{array}{cccccc}
1,&3/2&&&&\\
&&&&&\\
0,&0,&1/6,&1/2,&5/6,&2
\end{array}
\right)
\eqe



\section*{References}
\begin{thebibliography}{10}

\bibitem{kogaetal06}
J.~{Koga}, T.~Z. {Esirkepov}, and S.~V. {Bulanov}.
\newblock {Nonlinear Thomson scattering with strong radiation damping}.
\newblock {\em Journal of Plasma Physics}, 72:1315--+, December 2006.

\bibitem{muelleretal08}
C.~{M{\"u}ller}, A.~{di Piazza}, A.~{Shahbaz}, T.~J. {B{\"u}rvenich},
  J.~{Evers}, K.~Z. {Hatsagortsyan}, and C.~H. {Keitel}.
\newblock {High-energy, nuclear, and QED processes in strong laser fields}.
\newblock {\em Laser Physics}, 18:175--184, March 2008.

\bibitem{dipiazzahatsagortsyan08}
A.~{di Piazza} and K.~Z. {Hatsagortsyan}.
\newblock {Quantum interaction among intense laser beams in vacuum}.
\newblock {\em European Physical Journal Special Topics}, 160:147--155, July
  2008.

\bibitem{bellkirk08}
A.~R. {Bell} and J.~G. {Kirk}.
\newblock {Possibility of Prolific Pair Production with High-Power Lasers}.
\newblock {\em Physical Review Letters}, 101(20):200403--+, November 2008.

\bibitem{sheareretal73}
J.~W. {Shearer}, J.~{Garrison}, J.~{Wong}, and J.~E. {Swain}.
\newblock {Pair Production by Relativistic Electrons from an Intense Laser
  Focus}.
\newblock {\em Physical Review A}, 8:1582--1588, September 1973.

\bibitem{liangetal98}
E.~P. {Liang}, S.~C. {Wilks}, and M.~{Tabak}.
\newblock {Pair Production by Ultraintense Lasers}.
\newblock {\em Physical Review Letters}, 81:4887--4890, November 1998.

\bibitem{cowanetal99}
T.~E. {Cowan}, M.~D. {Perry}, M.~H. {Key}, T.~R. {Ditmire}, S.~P. {Hatchett},
  E.~A. {Henry}, J.~D. {Moody}, M.~J. {Moran}, D.~M. {Pennington}, T.~W.
  {Phillips}, T.~C. {Sangster}, J.~A. {Sefcik}, M.~S. {Singh}, R.~A. {Snavery},
  M.~A. {Stoyer}, S.~C. {Wilks}, P.~E. {Young}, Y.~{Takahashi}, B.~{Dong},
  W.~{Fountain}, T.~{Parnell}, J.~{Johnson}, A.~W. {Hunt}, and T.~{K\"uhl}.
\newblock {High energy electrons, nuclear phenomena and heating in petawatt
  laser-solid experiments}.
\newblock {\em Laser Part.\ Beams}, 17:773--783, October 1999.

\bibitem{nakashimaetal02}
K.~{Nakashima}, T.~E. {Cowan}, and H.~{Takabe}.
\newblock {Electron-Positron Pair Production by Ultra-Intense Lasers}.
\newblock In K.~{Nakajima} and M.~{Deguchi}, editors, {\em Science of
  Superstrong Field Interactions}, volume 634 of {\em American Institute of
  Physics Conference Series}, pages 323--328, October 2002.

\bibitem{chenetal09}
H.~{Chen}, S.~C. {Wilks}, J.~D. {Bonlie}, E.~P. {Liang}, J.~{Myatt}, D.~F.
  {Price}, D.~D. {Meyerhofer}, and P.~{Beiersdorfer}.
\newblock {Relativistic Positron Creation Using Ultraintense Short Pulse
  Lasers}.
\newblock {\em Physical Review Letters}, 102(10):105001--+, March 2009.

\bibitem{schwinger51}
J.~{Schwinger}.
\newblock {On Gauge Invariance and Vacuum Polarization}.
\newblock {\em Physical Review}, 82:664--679, June 1951.

\bibitem{bulanovetal06}
S.~S. {Bulanov}, N.~B. {Narozhny}, V.~D. {Mur}, and V.~S. {Popov}.
\newblock {Electron-positron pair production by electromagnetic pulses}.
\newblock {\em Soviet Journal of Experimental and Theoretical Physics},
  102:9--+, January 2006.

\bibitem{rufetal09}
M.~{Ruf}, G.~R. {Mocken}, C.~{M{\"u}ller}, K.~Z. {Hatsagortsyan}, and C.~H.
  {Keitel}.
\newblock {Pair production in laser fields oscillating in space and time}.
\newblock {\em Physical Review Letters}, 102:080402--+, 2009.

\bibitem{hebenstreitetal09}
F.~{Hebenstreit}, R.~{Alkofer}, G.~V. {Dunne}, and H.~{Gies}.
\newblock {Momentum signatures for Schwinger pair production in short laser
  pulses with a sub-cycle structure}.
\newblock {\em Physical Review Letters}, 102:150404--+, 2009.

\bibitem{burkeetal97}
D.~L. {Burke}, R.~C. {Field}, G.~{Horton-Smith}, J.~E. {Spencer}, D.~{Walz},
  S.~C. {Berridge}, W.~M. {Bugg}, K.~{Shmakov}, A.~W. {Weidemann}, C.~{Bula},
  K.~T. {McDonald}, E.~J. {Prebys}, C.~{Bamber}, S.~J. {Boege}, T.~{Koffas},
  T.~{Kotseroglou}, A.~C. {Melissinos}, D.~D. {Meyerhofer}, D.~A. {Reis}, and
  W.~{Ragg}.
\newblock {Positron Production in Multiphoton Light-by-Light Scattering}.
\newblock {\em Physical Review Letters}, 79:1626--1629, September 1997.

\bibitem{ritus79}
V.~I. {Ritus}.
\newblock {Quantum effects in the interaction of elementary particles with an
  intense electromagnetic field}.
\newblock {\em Moscow Izdatel Nauka AN SSR Fizicheskii Institut Trudy},
  111:5--151, 1979.

\bibitem{baierkatkov68}
V.~N. {Ba{\v i}er} and V.~M. {Katkov}.
\newblock {Processes Involved in the Motion of High Energy Particles in a
  Magnetic Field}.
\newblock {\em Soviet Journal of Experimental and Theoretical Physics},
  26:854--+, April 1968.

\bibitem{baevaetal06}
T.~{Baeva}, S.~{Gordienko}, and A.~{Pukhov}.
\newblock {Theory of high-order harmonic generation in relativistic laser
  interaction with overdense plasma}.
\newblock {\em Phys.\ Rev.\ E}, 74(4):046404--+, October 2006.

\bibitem{baeva08}
T.~{Baeva}.
\newblock {Coherent X-rays from relativistic plasmas}.
\newblock {\em Annual report of the UK Central Laser Facility}, pages 103 --
  106, 2008.

\bibitem{erber66}
T.~{Erber}.
\newblock {High-Energy Electromagnetic Conversion Processes in Intense Magnetic
  Fields}.
\newblock {\em Reviews of Modern Physics}, 38:626--659, October 1966.

\bibitem{melrose80}
D.~B. {Melrose}.
\newblock {\em {Plasma Astrohysics. Nonthermal processes in diffuse magnetized
  plasmas - Vol.1: The emission, absorption and transfer of waves in plasmas;
  Vol.2: Astrophysical applications}}.
\newblock New York: Gordon and Breach, 1980, 1980.

\bibitem{landaulifshitz75}
L.~D. {Landau} and E.~M. {Lifshitz}.
\newblock {\em {The Classical Theory of Fields}}.
\newblock Course of theoretical physics - Pergamon International Library of
  Science, Technology, Engineering and Social Studies, Oxford: Pergamon Press,
  1975, 4th rev.engl.ed., 1975.

\bibitem{klepikov85}
N.~P. {Klepikov}.
\newblock {Methodological Notes: Radiation damping forces and radiation from
  charged particles}.
\newblock {\em Soviet Physics Uspekhi}, 28:506--520, June 1985.

\bibitem{dipiazza08}
A.~{di~Piazza}.
\newblock {Exact solution of the Landau-Lifshitz equation}.
\newblock {\em Lett.\ Math.\ Phys.}, 83:305--313, 2008.

\bibitem{sokolovternov68}
A.~A. {Sokolov} and I.~M. {Ternov}.
\newblock {\em {Synchrotron Radiation}}.
\newblock Berlin: Akademie-Verlag, 1968.

\bibitem{akhiezeretal94}
A.~I. {Akhiezer}, N.~P. {Merenkov}, and A.~P. {Rekalo}.
\newblock {On a kinetic theory of electromagnetic showers in strong magnetic
  fields}.
\newblock {\em Journal of Physics G Nuclear Physics}, 20:1499--1514, September
  1994.

\end{thebibliography}

\end{document}